\documentclass[10pt, prb, aps, twocolumn, showpacs, citeautoscript, floatfix, reprint, amsmath, amssymb, notitlepage]{revtex4-1}
\usepackage{booktabs}
\usepackage[utf8]{inputenc}
\usepackage[T1]{fontenc}
\usepackage[english]{babel}
\usepackage{graphicx}
\usepackage{rotating}
\usepackage{mathtools}
\usepackage{amsfonts}
\usepackage{dcolumn} % align table columns on decimal point
\usepackage{bm} % bold math
\usepackage{xcolor}
\usepackage{latexsym}
\usepackage{wasysym}
\usepackage[colorlinks, citecolor=blue, linkcolor=blue, urlcolor=blue]{hyperref}
\usepackage[nomain, acronym, shortcuts]{glossaries}
\usepackage{tikz}
\usepackage{environ}
\usepackage{csquotes}
\glsdisablehyper
\usetikzlibrary{external}
\definecolor{blue}{rgb}{0,0,.5}

\bibpunct{[}{]}{,}{n}{}{}

\newcommand{\commabr}{,\allowbreak}

\DeclareMathOperator{\sgn}{sgn}

\newcommand{\eq}[1]{Eq.~\eqref{#1}}
\newcommand{\bs}[1]{{\bm{#1}}}
\newcommand{\bk}{\bs{k}}
\newcommand{\br}{\bs{r}}
\newcommand{\bR}{\bs{R}}
\newcommand{\bd}{\bs{d}}
\newcommand{\bsigma}{\bs{\sigma}}

\renewcommand{\i}{\mathrm{i}}
\renewcommand*\d{\mathop{}\!\mathrm{d}}
\newcommand{\e}[1]{\mathrm{e}^{#1}}
\newcommand{\rmc}{\mathrm{c}}
\newcommand{\hc}{\text{H.c.}}
\newcommand{\td}{t_{d}}
\newcommand{\tdd}{t_{d}'}
\newcommand{\tf}{t_{f}}
\newcommand{\tff}{t_{f}'}
\newcommand{\ef}{\epsilon_{f}}
\newcommand{\Nn}[1]{\langle #1\rangle}
\newcommand{\Nnn}[1]{\langle\!\langle #1 \rangle\!\rangle}
\newcommand{\transp}{\mathrm{T}}
\newcommand{\xtimesx}[1]{$#1{\times}#1$}
\newcommand{\const}{\text{const}}
\newcommand{\nodag}{\vphantom{\dag}}
\newcommand{\norm}[1]{\left|\!\left|#1\right|\!\right|}

\newacronym{hsp}{HSP}{high-symmetry point}
\newacronym{ssh}{SSH}{Su--Schrieffer--Heeger}
\newacronym{bhz}{BHZ}{Bernevig-Hughes-Zhang}
\newacronym{ti}{TI}{topological insulator}
\newacronym{tki}{TKI}{topological Kondo insulator}
\newacronym{tsc}{TSC}{topological superconductor}
\newacronym{rg}{RG}{renormalization-group}
\newacronym{mcn}{MCN}{mirror Chern number}
\newacronym{bz}{BZ}{Brillouin zone}

\NewEnviron{mysplit}{%
	\begin{equation}
	\begin{split}
	\BODY
	\end{split}
	\end{equation}
}

\graphicspath{{Figures/}}

\begin{document}
\newcommand{\mytitle}{Correlation length, universality classes, and scaling laws associated with 
topological phase transitions}
\title{\mytitle}

\newcommand{\affeth}{\affiliation{Institut f\"ur Theoretische Physik, ETH Z\"urich, 8093 Z\"urich, Switzerland}}

\author{Wei Chen}\affeth
\author{Markus Legner}\affeth
\author{Andreas R\"{u}egg}\affeth  
\author{Manfred Sigrist}\affeth

\date{\today}

\hypersetup{
	pdftitle={\mytitle}
}

\begin{abstract}
%\notem{Maybe an additional introductory sentence?}
The correlation functions related to topological phase transitions in inversion-symmetric lattice models described by \xtimesx{2} Dirac Hamiltonians are discussed.
In one dimension, the correlation function measures the charge-polarization correlation between Wannier states at different positions, while in two dimensions it measures the itinerant-circulation correlation between Wannier states.
The correlation function is nonzero in both the topologically trivial and nontrivial states, and allows to extract a correlation length that diverges at topological phase transitions.
The correlation length and the curvature function that defines the topological invariants are shown to have universal critical exponents, allowing the notion of universality classes to be introduced.
Particularly in two dimensions, the universality class is determined by the orbital symmetry of the Dirac model.
The scaling laws that constrain the critical exponents are revealed, and are predicted to be satisfied even in interacting systems, as demonstrated in an interacting \acrlong*{tki}.
\end{abstract}


%74.20.De Phenomenological theories (two-fluid, Ginzburg-Landau, etc.)
%64.60.ae Renormalization-group theory
%64.60.F- Equilibrium properties near critical points, critical exponents
%73.20.-r Electron states at surfaces and interfaces

\maketitle

\section{Introduction}

A central issue in the study of \acp{ti} and \acp{tsc} is how to drive the system into the topologically nontrivial state such that their intriguing properties, for instance the chiral edge currents~\cite{Haldane88,Kane05,Kane05_2,Bernevig06,Konig07} and Majorana edge states~\cite{Kitaev01,Lutchyn10,Oreg10,Mourik12}, can be exploited.
The topologically nontrivial state is characterized by a nonzero topological invariant that is often calculated from the integration of certain curvature function such as Berry connection~\cite{Zak89} or Berry curvature~\cite{Berry84,Xiao10,Thouless82}, while in the trivial state the topological invariant is zero.
The transition from a topologically trivial to a nontrivial state necessarily involves inverting the bulk bands, which is usually achieved by tuning an energy parameter that may be the chemical potential~\cite{Kitaev01,Lutchyn10,Oreg10,Mourik12}, hopping amplitudes~\cite{Su79}, interface coupling~\cite{Choy11,Nadj-Perge13,Braunecker13,Pientka13,Klinovaja13,Vazifeh13,Rontynen14,Kim14,Sedlmayr15,Chen15_Majorana}, etc. In this paper, we will refer to this parameter as the \enquote{tuning energy parameter}.
Consequently, topological phase transitions can be identified by calculating the values of the tuning energy parameter at which the bulk gap closes, provided that topologically nontrivial states can exist according to the symmetry classification~\cite{Schnyder08,Kitaev09}.

On the other hand, the concept of scaling and scale invariance has recently emerged as an alternative way to judge topological phase transitions~\cite{Chen16,Chen16_2}.
Akin to stretching a messy string to reveal the number of knots it contains, the scaling procedure locally renormalizes a curvature function without changing the topological invariant.
In this scheme, the  \ac{rg} flow uncovers the fix points of the curvature function of the topologically trivial and non-trivial phases.

In this article, we further advance such an approach to topological phase transitions by answering the following fundamental questions, focusing on 1D and 2D inversion-symmetric lattice models described by \xtimesx{2} Dirac Hamiltonians, where the bulk gap closes at \acp{hsp}~\cite{Murakami11}: 
(1) Since topological systems may or may not possess an order parameter from which a correlation function is usually defined, can there be a correlation function that is ubiquitous in \acp{ti}?
(2) If such a correlation function exists, how is the correlation length related to other fundamental length scales in the problem, especially the edge-state decay length~\cite{Konig08,Zhou08,Linder09,Lu10,Shen12,Qi11} and the length scale that characterizes the scale invariance at the critical point~\cite{Chen16,Chen16_2}? 
(3) As topological phase transitions can be driven by all kinds of tuning energy parameters that invert the bulk bands, can there be universal critical exponents and scaling laws regardless of what the tuning energy parameter is?
(4) Do different sets of critical exponents correspond to different universality classes and, if so, what is the underlying symmetry that determines the universality class?

We show that these delicately intertwined issues can be addressed in a unified manner using the Wannier-state representation~\cite{KingSmith93,Resta94,Resta07} in combination with \xtimesx{2} Dirac Hamiltonians.
We identify the sought-after correlation function in 1D as the Fourier transform of the Berry connection, which yields a filled-band contribution to the charge-polarization correlation between Wannier states at different positions, a quantity that has been discussed within the theory of charge polarization~\cite{KingSmith93,Resta94,Resta07,Bernevig13,Xiao10}.
The formalism extended to 2D renders a filled-band Wannier-state correlation function calculated from the Fourier transform of the Berry curvature, which is intimately related to the itinerant-circulation part of the orbital magnetization~\cite{Thonhauser05,Xiao05,Ceresoli06,Shi07,Souza08}.
Remarkably, we find that these Wannier-state correlation functions are nonzero in both topologically trivial and nontrivial states.
Close to the critical point, the correlation length essentially coincides with the diverging length scale suggested by the scaling scheme, and in the topologically nontrivial state it coincides with the edge-state decay length.
In addition, since the tuning energy parameter is a scalar that must enter the even-parity channel of the Dirac Hamiltonian, the curvature function and the correlation length of noninteracting Dirac Hamiltonians display universal critical exponents regardless of the tuning energy parameter.
Particularly in 2D, the critical exponents can be classified into different universality classes according to the orbital symmetry of the Dirac Hamiltonian.
The scaling laws that constrain the critical exponents are identified, and are predicted to be satisfied even in interacting systems. This is demonstrated by model calculations for an interacting \ac{tki} formulated within a slave-boson-mean-field theory~\cite{Legner14}.

The remainder of the article is arranged in the following way.
Section~\ref{1D_formalism} discusses the quantities of interest in 1D noninteracting lattice models, including correlation function, correlation length and its correspondence to the edge-state decay length, scaling laws, and the application to a generic \xtimesx{2} Dirac Hamiltonian.
Section~\ref{Sec_critical_exponents} generalizes the formalism to 2D systems and elaborates on the relation between the universality class and the orbital symmetry.
The interacting \ac{tki} is formulated in Sec.~\ref{sec:tki}.
In Sec.~\ref{sec:concl}, we summarize the results and discuss the common features in all these models.

\section{1D noninteracting systems \label{1D_formalism}}

\subsection{Correlation length and scaling laws in 1D noninteracting systems \label{1D_scaling_laws}}

Consider a $D$-dimensional inversion-symmetric lattice model whose topology is described by the topological invariant 
%\notem{we only explicitly write $\d^D \bk$ here, everywhere else it's just $\d \bk$}
\begin{align}
\mathcal{C}=\int\frac{\d\bk}{\left(2\pi\right)^{D}}F(\bk,M)\;,
\label{1D_winding_number}
\end{align} 
where $F(\bk,M)$ is referred to as the curvature function at momentum $\bk$ in $D$ dimensions.
For $D=1$, the curvature function is the Berry connection~\cite{Zak89,Xiao10} summed over occupied bands $n$,
\begin{align}
F(k,M)=\sum_{n}\langle u_{nk}|\i \partial_{k}|u_{nk}\rangle\;,
\label{1D_Berry_connection}
\end{align}
where $|u_{nk}\rangle$ is the Bloch state at momentum $k$ with band index $n$.
Throughout the article, we use $M$ as the tuning energy parameter that controls the topology of the system.
The Berry connection is gauge dependent~\cite{Xiao10} and changes under $|u_{nk}\rangle\rightarrow \e{\i \eta_{nk}}|u_{nk}\rangle$.
In the following discussion for 1D inversion-symmetric lattice models, we choose the gauge that gives a Berry connection that has a Lorentzian shape around the gap-closing momentum $k_{0}$
\begin{align}
F(k_{0}+\delta k,M)=\frac{F(k_{0},M)}{1\pm\xi^{2}\delta k^{2}}\;,
\label{xi_definition}
\end{align}
where $\xi$ is a characteristic length scale~\cite{Chen16,Chen16_2}.
It is in this gauge choice that the previously proposed scaling scheme is applicable in 1D systems since it relies on the form of \eq{xi_definition}.
We will demonstrate explicitly this gauge choice for Dirac models. %{\color{green} Is this gauge choice related to the gauge which produces maximally localized Wannier funtions? If so, we should state it here.}

The length scale $\xi$ in \eq{xi_definition} has been suggested to characterize the scale invariance at the critical point~\cite{Chen16} since it diverges there.
As \eq{xi_definition} clearly resembles the Ornstein--Zernike form of a correlation function in momentum space, we now demonstrate that $\xi$ indeed represents a correlation length in 1D.
Consider the Wannier states for $D$-dimensional lattice models
\begin{subequations}
\begin{align}
|u_{n\bk}\rangle&=\sum_{\bR}\e{-\i \bk\cdot(\br-\bR)}|\bR n\rangle\;,
 \\
|\bR n\rangle&=\int \d \bk\,\e{\i \bk\cdot(\br-\bR)}|u_{n\bk}\rangle\;,
\end{align}\label{Wannier_basis}%
\end{subequations} 
and their wave functions $\langle r|\bR n\rangle=W_{n}(\br-\bR)$.
Particularly in $D=1$, the charge polarization 
\begin{equation}
\mathcal{P}=\sum_{n}\langle 0n|r|0n\rangle=\int \d k\sum_{n}\langle u_{nk}|\i\partial_{k}|u_{nk}\rangle
\label{charge_polarization}
\end{equation}
is known to be equal to the topological invariant in \eq{1D_winding_number}, up to an integer under gauge transformation~\cite{KingSmith93,Resta94,Resta07,Bernevig13,Xiao10}, where $n$ sums over filled bands.
Moreover, in the Wannier-state representation, the contribution to the Berry connection from the $n$\textsuperscript{th} filled band is~\cite{Marzari97,Marzari12,Gradhand12}
\begin{align}
\begin{split}
\langle u_{nk}|\i\partial_{k}|u_{nk}\rangle&=\sum_{R}\e{\i kR}\langle 0n|r|Rn\rangle
\\
&=\sum_{R}\e{-\i kR}\langle Rn|r|0n\rangle\;,
\end{split}
\end{align}
from which it follows that the Fourier component of the curvature function in \eq{1D_Berry_connection} takes the form~\cite{Marzari97,Wang06}
\begin{mysplit}
\lambda_{R}&=\int \d k\; \e{\i kR}F(k,M)=\sum_{n}\langle Rn|r|0n\rangle
 \\
&=\sum_{n}\int r\;W_{n}(r-R)^{\ast}W_{n}(r)\d r\;,
\label{1D_Fourier_component}
\end{mysplit}
which sums the diagonal matrix elements of the position operator between Wannier states in the filled bands.
Using~\cite{Bernevig13} $\langle u_{nk}|\i \partial_{k}|u_{nk}\rangle=\left(-\i \partial_{k}\langle u_{nk}|\right)|u_{nk}\rangle$ and inversion symmetry, one sees that $\lambda_{R}^{\ast}=\lambda_{-R}=\lambda_{R}$ is a real function. We observe that in the continuous limit, the Fourier transform of the Ornstein--Zernike form of \eq{xi_definition} yields
\begin{align}
\left[1\mp\xi^{2}\partial_{R}^{2}+f(\partial_{R})\right]\lambda_{R}=F(k_{0},M)\delta(R)\;,
\label{1D_lambda_equation}
\end{align}
where $f(\partial_{R})$ is a polynomial of $\partial_{R}$ that represents the structures other than the singularity at $k_{0}$.
This suggests that $\lambda_{R}$ to leading order is a function of $R/\xi$,
\begin{align}
\lambda_{R}=\lambda(R/\xi)\;.
\label{lambdar_1D}
\end{align}
Since $\lambda_{R}$ which can be viewed as a charge-polarization correlation function between the Wannier states $|0n\rangle$ and $|Rn\rangle$ summing over filled bands, $\xi$ unambiguously plays the role of a correlation length. In addition, the onsite correlation $\lambda_{0}$ coincides with the charge polarization in Eq.~\eqref{charge_polarization} and hence the topological invariant. 
Notice that, however, if one chooses a different gauge $|u_{nk}\rangle\rightarrow \e{\i \eta_{nk}}|u_{nk}\rangle$ in which the Berry connection does not have the Ornstein--Zernike form in \eq{xi_definition}, then the correlation function is modified accordingly $\lambda_{R}\rightarrow\lambda_{R}-\sum_{n}\int dk\;\e{\i kR}\;\partial_{k}\eta_{nk}$ and in general will not be a function that simply decays with $\xi$. 
Thus, our discussion in 1D is limited to a specific gauge.

\begin{figure}
\centering
\includegraphics[clip=true,width=0.99\columnwidth]{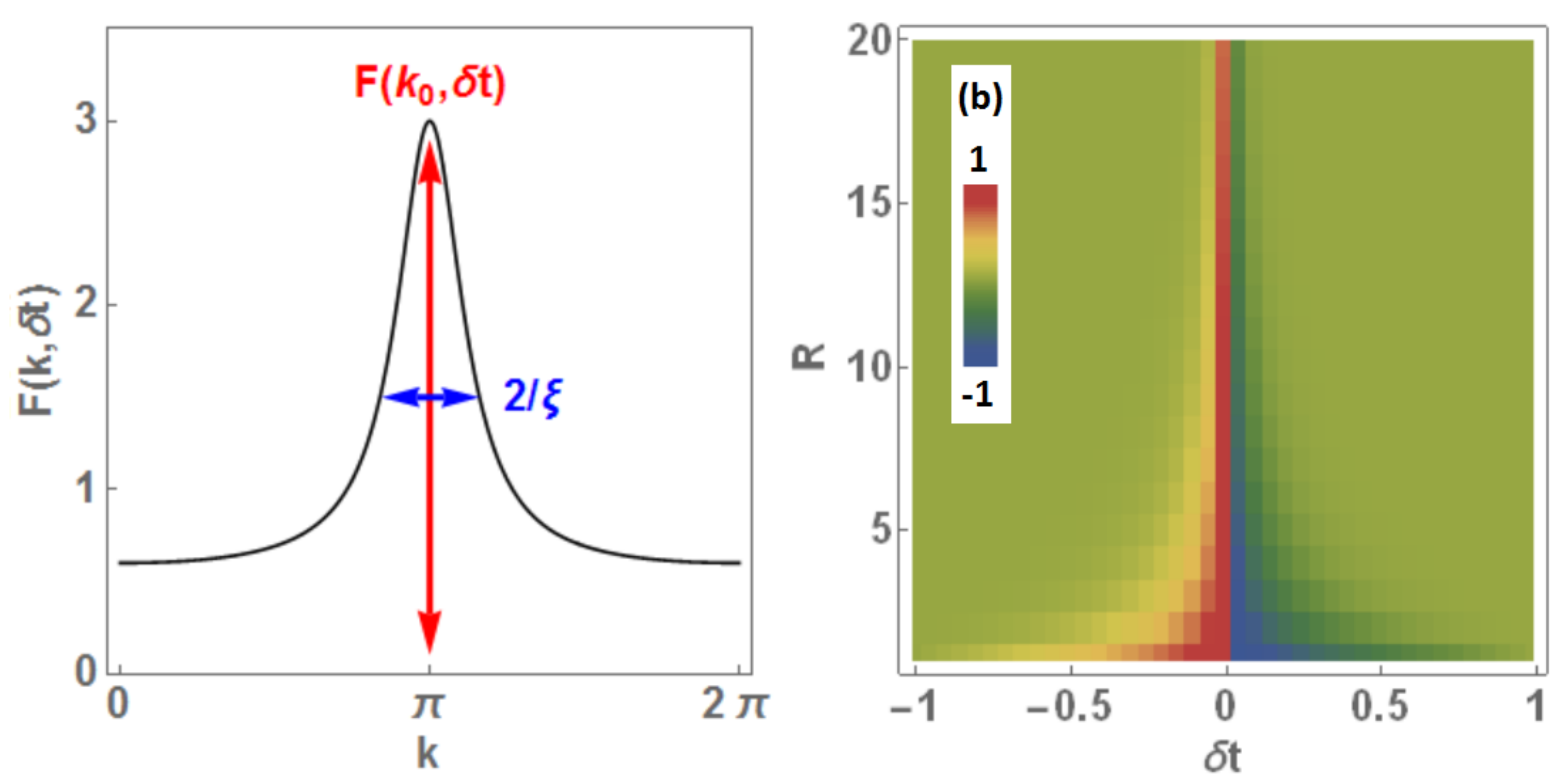}
\includegraphics[clip=true,width=0.95\columnwidth]{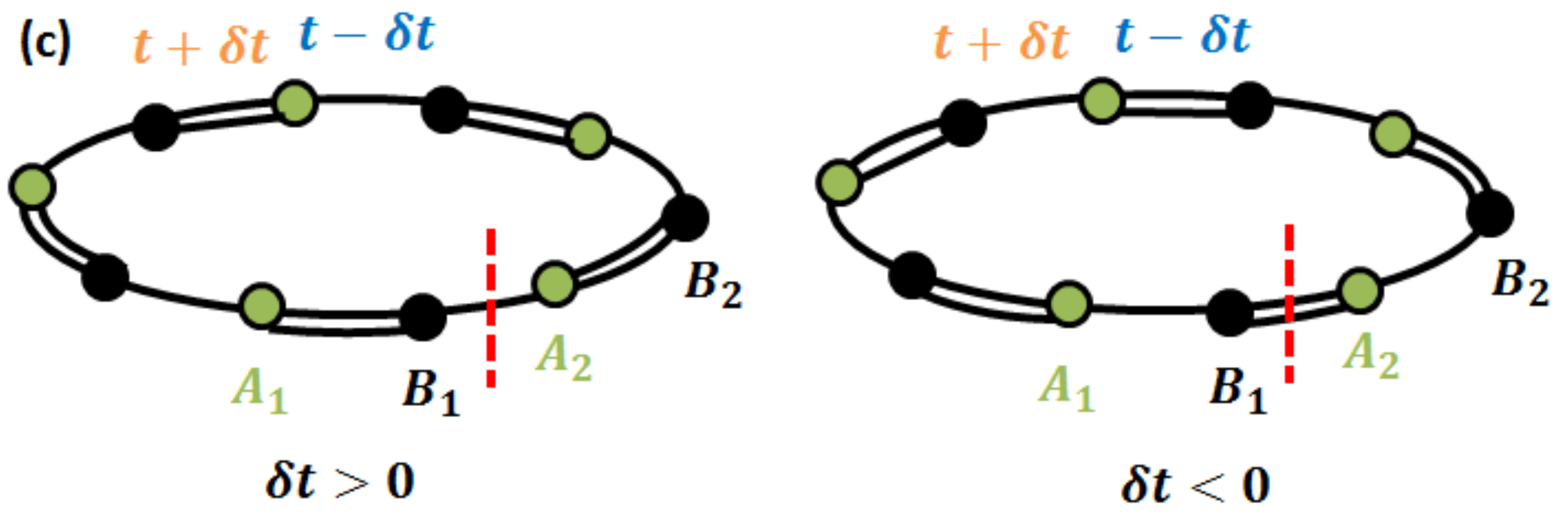}
\caption{(color online) (a) The curvature function (Berry connection within the described gauge choice) $F(k,\delta t)$ of the \ac{ssh} model at $\delta t/t=-0.2$, which peaks at the gap-closing momentum $k_{0}=\pi$.
The half width at half maximum extracts a length scale $\xi$ that is the correlation length in \eq{lambdar_1D}.
The area under the singular peak converges to a constant as $\delta t\rightarrow\delta t_\rmc$, as described by $\mathcal{C}_\mathrm{div}$ in \eq{1D_Cprime}.
(b) The Fourier component $2\lambda_{R}$ of $F(k,\delta t)$ described by \eq{SSH_lambdar}, which also represents the Wannier-state charge-polarization correlation function described in \eq{1D_Fourier_component}.
It exponentially decays in both $R$ and $\delta t-\delta t_\rmc$ and changes sign at $\delta t_\rmc$.
(c) Schematics for the topologically trivial phase $\delta t>0$ (left) which dimerizes between $\mathcal A_{i}$ and $\mathcal B_{i}$ and has no edge state when the ring is cut at the dashed line, and the topologically nontrivial phase $\delta t<0$ (right) that dimerizes between $\mathcal B_{i}$ and $\mathcal A_{i+1}$ and has edge state when the ring is cut at the dashed line.
The Wannier-state charge-polarization correlation for the two cases are however the same since it only depends on the relative distance between two Wannier home cells.} 
\label{fig:SSH_model_lambda_exponents}
\end{figure}

We proceed to discuss a scaling law in 1D.
As $M\rightarrow M_\rmc$, the curvature function at the gap-closing momentum $k_{0}$ diverges, yet its integration over the range of $\xi^{-1}$ near $k_{0}$ converges to a constant (see Fig.~\ref{fig:SSH_model_lambda_exponents} (a) for demonstration):
\begin{align}
\begin{split}
\mathcal{C}_\mathrm{div}&=\int_{-\xi^{-1}}^{\xi^{-1}}\frac{\d k}{2\pi}F(k,M)
=\int_{-\xi^{-1}}^{\xi^{-1}}\frac{\d k}{2\pi}\frac{F(k_{0},M)}{1\pm\xi^{2}k^{2}}
 \\
&=\frac{F(k_{0},M)}{\xi}\times\mathcal{O}(1)=\const\;.
\end{split}
\label{1D_Cprime}
\end{align}   
Thus, $F(k_{0},M)$ and $\xi$ have the same critical exponent in 1D systems:
\begin{subequations}
\begin{align}
|F(k_{0},M)|&\propto|M-M_\rmc|^{-\gamma}\;,
 \\
\xi&\propto|M-M_\rmc|^{-\nu}\;,
 \\
\gamma&=\nu\;.
\label{1D_critical_exponent_summary}
\end{align}
\end{subequations}
Note that since $F(k_{0},M)$ is the correlation function integrated over space, $F(k_{0},M)=\sum_{R}\lambda_{R}$, it plays a role similar to the susceptibility in the Landau-order-parameter paradigm.
For this reason, we denote its critical exponent by $-\gamma$ following the usual convention for susceptibility.
As we shall see in the example below, as $M\rightarrow M_\rmc$, one may parametrize $F(k_{0},M)$ by
% \begin{subequations}
\begin{align}
F(k_{0},M)
&\propto \sgn(M-M_\rmc)|M-M_\rmc|^{-\gamma}
\;,
%  \\
% \sgn(\eta)|_{M>M_\rmc}&=-\sgn(\eta)|_{M<M_\rmc}\;.
\label{sign_of_eta}
\end{align}
% \end{subequations}
since $F(k_{0},M)$ changes sign between $M<M_\rmc$ and $M>M_\rmc$.

\subsection{Critical exponents of 1D \texorpdfstring{2$\bm{\times}$2}{2x2} Dirac Hamiltonians of symmetry class AIII \label{1D_2by2_section}}

To further demonstrate that noninteracting 1D Dirac models exhibit universal critical exponents, we consider a generic 1D spinless inversion-symmetric model of symmetry class AIII described by a \xtimesx{2} Dirac Hamiltonian
\begin{align}
\label{eq:dirac_1d}
H(k)=\sum_{i}d_{i}(k)\sigma_{i}\;,
\end{align}
where the two degrees of freedom $\psi=\left(c_{\mathcal A},c_{\mathcal B}\right)^\transp$ come from $\mathcal{A}$ and $\mathcal{B}$ sublattices.
The chiral (sublattice) symmetry of class AIII demands $\sigma_{z}H(k)+H(k)\sigma_{z}=0$ and therefore $d_{3}=0$.
The gauge choice that gives \eq{xi_definition} can be constructed in the following way.
The eigenstates with eigenenergies $E_{\pm}=\pm d=\pm\sqrt{d_{1}^{2}+d_{2}^{2}}$ are chosen to be in the gauge
\begin{align}
|\psi_{\pm}\rangle=\frac{1}{\sqrt{2}d}
\begin{pmatrix}
\pm d \\
d_{1}+\i d_{2}
\end{pmatrix}
\;.
\label{1D_2by2_wave_function}
\end{align}
The Berry connection constructed from the filled band in this gauge choice is
\begin{align}
A_{k}=\i \langle\psi_{-}|\partial_{k}|\psi_{-}\rangle=\frac{d_{2}\partial_{k}d_{1}-d_{1}\partial_{k}d_{2}}{2d^{2}}=F(k,M)\;.
\label{Berry_connection_general}
\end{align}
The Dirac Hamiltonian near the gap-closing momentum $k-k_{0}=\delta k\rightarrow 0$ and close to the critical point $M\rightarrow M_\rmc$ is generically parametrized by either $(d_{1},d_{2})=(\mathrm{even},\mathrm{odd})$ in $\delta k$ or $(d_{1},d_{2})=(\mathrm{odd},\mathrm{even})$ in $\delta k$, depending on the model.
For a model described by the first case $(d_{1},d_{2})=(\mathrm{even},\mathrm{odd})$, an expansion to leading order yields
\begin{subequations}
\label{1D_2by2_di_general}
\begin{align}
d_{1}&=(M-M_\rmc)+B\delta k^{2}\;,
 \\
d_{2}&=A\delta k\;,
 \\
d_{3}&=0\;.
\end{align}
\end{subequations}
The band-inversion parameter $M$, which drives the topological phase transition, has to enter the parity-even $d_{1}$ component because it is a scalar.
For the second case $(d_{1},d_{2})=(\mathrm{odd},\mathrm{even})$, one swaps the functional form for $d_{1}$ and $d_{2}$ in \eq{1D_2by2_di_general}, which merely gives an overall minus sign to the Berry connection in \eq{Berry_connection_general}, so the analysis of critical exponents below remains valid.

\begin{figure}
\centering
\includegraphics[clip=true,width=0.99\columnwidth]{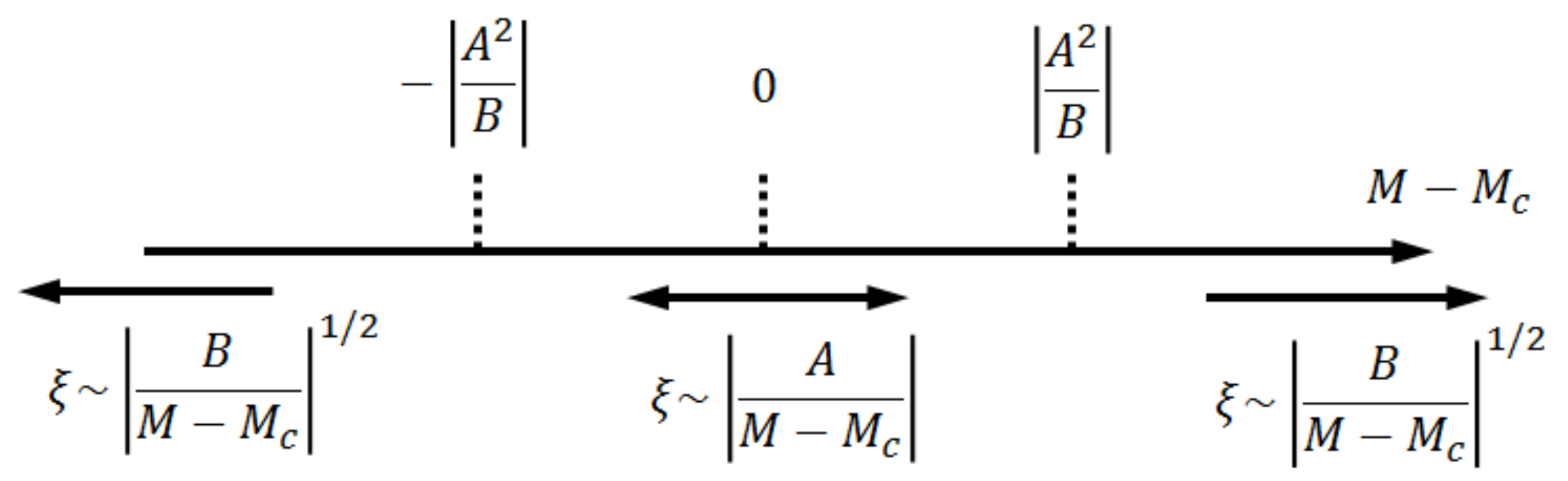}
\caption{Correlation length $\xi$ versus tuning energy parameter $M-M_\rmc$ in a generic Dirac Hamiltonian parametrized by the velocity $A$ and inverse effective mass $B$ as in Eqs.~\eqref{1D_2by2_di_general} and \eqref{di_general_form}.
Inside the critical region $|M-M_\rmc|<|A^{2}/B|$ the correlation length is approximately $\xi\sim|A/(M-M_\rmc)|$, while outside the critical region it is approximately $\xi\sim|B/(M-M_\rmc)|^{1/2}$.} 
\label{fig:xi_versus_Gamma}
\end{figure}

Putting \eq{1D_2by2_di_general} into \eq{xi_definition} yields a Berry connection that is symmetric around $k_{0}$ (this is the purpose of the gauge choice in \eq{1D_2by2_wave_function}), whose expansion gives 
\begin{subequations}
\label{1D_2by2_A0_xi}
\begin{align}
A_{k_{0}+\delta k}&=\frac{A_{k_{0}}}{1\pm\xi^{2}\delta k^{2}}\;,
 \\
A_{k_{0}}&=-\frac{A}{2(M-M_\rmc)}\propto(M-M_\rmc)^{-1}\;,
 \\
 \begin{split}
\xi&=\left|
\frac{3B(M-M_\rmc)+A^{2}}{(M-M_\rmc)^{2}}\right|^{1/2}\\
&\overset{\mathclap{M\to M_\rmc}}{\propto}\quad |M-M_\rmc|^{-1}\;.
 \end{split}
\end{align}
\end{subequations}
%{\color{green}(Here and below \newm{and in Fig.~1}, you are using the parameter $A$, while in \eq{1D_2by2_di_general} you are using $A_2$. What is the connection?)} 
One sees that as $M\rightarrow M_\rmc$, $\xi$ is always real and positive regardless of the precise values of $(B,A)$.
The critical exponent of $A_{k_{0}}$ and of $\xi$ are $\gamma=\nu=1$, in accordance with the scaling law described by \eq{1D_critical_exponent_summary}.

Two distinct features of the correlation length should be emphasized.
Firstly, out of the general parametrization in \eq{1D_2by2_di_general}, there are three fundamental length scales $A/(M-M_\rmc)$, $B/A$, and $|B/(M-M_\rmc)|^{1/2}$ embedded in the Hamiltonian.
As shown in Fig.~\ref{fig:xi_versus_Gamma}, as the tuning energy parameter approaches the critical point $M-M_\rmc=0$, there exists a critical region $|M-M_\rmc|<|A^{2}/B|$, inside of which the correlation length is dominated by $\xi\sim|A/(M-M_\rmc)|$;
outside of the critical region, $|M-M_\rmc|>|A^{2}/B|$, the dominating length scale is $\xi\sim|B/(M-M_\rmc)|^{1/2}$.
Near the boundary of the critical region, the correlation length and the three length scales are all of the same order of magnitude.
Secondly, in Appendix~\ref{appendix_xi_decay} we explicitly elaborate on the correspondence between the correlation length within this gauge choice and another fundamental length scale in the problem, namely the decay length of the edge state in the topologically nontrivial phase.
Despite this correspondence, we emphasize that the Wannier-state correlation function is finite in both the topologically trivial and nontrivial state, while the edge state only manifests at the interface between topologically distinct states.
%\notem{As it is stated now, this statement is not completely logical: Both the correlation function and the edge-state decay length are finite in both topological and trivial phase as long as there is an interface. The advantage of the correlation function is that it is also defined in the \emph{absence} of an interface.}

\subsection{Su--Schrieffer--Heeger model}

We proceed by discussing the \acrfull*{ssh} model~\cite{Su79} as a specific example to demonstrate these statements.
The class-AIII model describes spinless fermions on a closed ring with unequal hopping amplitudes on the even and odd bonds, 
\begin{align}
H&=\sum_{i}(t+\delta t)c_{\mathcal{A}i}^{\dag}c_{\mathcal{B}i}^{\nodag}+(t-\delta t)c_{\mathcal{A}i+1}^{\dag}c_{\mathcal{B}i}^{\nodag}+\hc \;,
\label{SSH_Hamiltonian}
\end{align}
where $\mathcal{A}$ and $\mathcal{B}$ are sublattice indices and $i$ is the site index.
The model has a spectral gap that closes at $k_{0}=\pi$, near which the Hamiltonian takes the form of \eq{1D_2by2_di_general} with 
\begin{subequations}
\begin{gather}
M-M_\rmc=2\delta t\;,\quad B=\frac{t-\delta t}{2}\;,\;
A=\delta t-t\;.
\end{gather}
\end{subequations}
The topological phase transition is driven by the difference in hopping amplitude $\delta t$.

Following the gauge choice of \eq{1D_2by2_wave_function}, we obtain the Berry connection for the SSH model as, with an additional factor of $-2$ for the sake of normalization, 
%{\color{green}(Although the prefactor should be unimportant for the scaling scheme, the $-2$ seems to appear from nowhere at this point)} 
\begin{align}
\begin{split}
F(k,\delta t)&=-2\i \langle\psi_{-}|\partial_{k}|\psi_{-}\rangle
 \\
&=\frac{\left(t^{2}-\delta t^{2}\right)\cos k+\left(t-\delta t\right)^{2}}{\left(t+\delta t\right)^{2}+2\left(t^{2}-\delta t^{2}\right)\cos k+\left(t-\delta t\right)^{2}}\;.
\end{split}
\label{SSH_Fk}
\end{align}
The topological invariant $\mathcal{C}$ can be calculated by a contour integration
\begin{align}
\begin{split}
\mathcal{C}&=\frac{1}{2\pi}\int_{0}^{2\pi} \d k \;F(k,\delta t)=\frac{1}{2}\left(1-\frac{t\;\delta t}{|t\;\delta t|}\right)
 \\
&=
\begin{cases}
0&\text{for }t\;\delta t>0\;, \\
1&\text{for }t\;\delta t<0\;,
\end{cases}
\end{split}
\label{Chern_number_SSH}
\end{align}
so the critical point is located at $\delta t_\rmc=0$.
At the gap-closing point $k_{0}=\pi$, \eq{SSH_Fk} gives
\begin{align}
\begin{split}
F(k_{0},\delta t)&=\frac{1}{2}\left(1-\frac{t}{\delta t}\right)\\
&\overset{\mathclap{\delta t\to \delta t_\rmc}}{\propto}\quad -\sgn\left(\frac{\delta t}{t}\right)|\delta t-\delta t_\rmc|^{-1}\;.
\end{split}
\label{SSH_Fk0}
\end{align}
The length scale $\xi$ defined from \eq{xi_definition} is~\cite{Chen16} 
\begin{align}
\xi&=\left|\frac{t}{4\delta t}\left(1+\frac{t}{\delta t}\right)\right|^{1/2}
\overset{{\delta t\to \delta t_\rmc}}{\propto} |\delta t-\delta t_\rmc|^{-1}\;.
\label{SSH_xi}
\end{align}
The Wannier-state charge-polarization correlation function is, using a contour integration for the Fourier component with the gap-closing momentum $k_{0}=\pi$ as the origin,
\begin{align}
\begin{split}
\lambda_{R}&=\int\frac{\d k}{2\pi}F(k,\delta t)\e{\i (k-\pi)R}\\
&\overset{\mathclap{\delta t\to \delta t_\rmc}}{\propto}\quad-\frac{1}{2}\sgn\left(\frac{\delta t}{t}\right)\e{-R/\xi}\;,
\end{split}
\label{SSH_lambdar}
\end{align}
which satisfies Eqs.~\eqref{lambdar_1D} and \eqref{sign_of_eta}, with $\xi$ playing the role of a correlation length.
Equations~\eqref{SSH_Fk0} and \eqref{SSH_xi} confirm that the \ac{ssh} model indeed satisfies the scaling law in \eq{1D_critical_exponent_summary} and the critical exponents in \eq{1D_2by2_A0_xi}.
Using Appendix~\ref{appendix_xi_decay}, the edge-state decay length is
\begin{align}
\xi_{-}\approx-\frac{t-\delta t}{2\delta t}\approx \xi\;.
\label{SSH_edge_decay_length}
\end{align}
Therefore, in the topologically nontrivial state and close to the critical point $\delta t/t\apprle 0$, the correlation length coincides with the decay length of edge states. Note, however, that the Wannier-state correlation is nonzero in both the topologically trivial and nontrivial state, as explained schematically in Fig.~\ref{fig:SSH_model_lambda_exponents} (c).

\section{2D noninteracting systems \label{Sec_critical_exponents}}

\subsection{Correlation length and scaling laws in 2D noninteracting systems}

For 2D systems with rectangular symmetry, we start from the curvature function that has the following form~\cite{Chen16,Chen16_2} near the gap-closing momentum $\bk_{0}$:
\begin{align}
F(\bk_{0}+\delta\bk,M)\approx\frac{F(\bk_{0},M)}{\left(1\pm\xi_{x}^{2}\delta k_{x}^{2}\right)\left(1\pm\xi_{y}^{2}\delta k_{y}^{2}\right)}\;.
\label{2D_xi_definition}
\end{align}
Depending on the scheme one chooses to calculate the topological invariant, $F(\bk,M)$ can be either the Berry curvature~\cite{Chen16}, the phase gradient of the Pfaffian of the time-reversal operator, or the second derivative of the Pfaffian~\cite{Chen16_2}.
A significant difference from the 1D case described in Sec.~\ref{1D_formalism}, is that these quantities are gauge-invariant in 2D, so there is no need to choose a particular gauge.

In the scheme that uses the filled-band Berry curvature as the curvature function,
\begin{subequations}
\begin{align}
A_{nk_{\alpha}}&=\langle u_{n\bk}|\i\partial_{k_{\alpha}}|u_{n\bk}\rangle\;,
\\
F(\bk,M)&=\sum_{n}\partial_{k_{x}}A_{nk_{y}}-\partial_{k_{y}}A_{nk_{x}}\;,
\end{align}
\end{subequations}
the length scale $\xi_{\alpha}$ in \eq{2D_xi_definition} again represents a correlation length. This can be seen by noticing that the $n$\textsuperscript{th} band contribution to the Berry curvature can be expressed in terms of the Wannier states in \eq{Wannier_basis} as~\cite{Wang06,Marzari12,Gradhand12},
\begin{align}
\begin{split}
\partial_{k_{x}}A_{nk_{y}}-\partial_{k_{y}}A_{nk_{x}}&=\i\sum_{\bR}\e{\i\bk\cdot\bR}\langle\bs{0} 
n|\left(\bR\times\br\right)_{z}|\bR n\rangle
\\
&=-\i\sum_{\bR}\e{-\i \bk\cdot\bR}\langle\bR n|\left(\bR\times\br\right)_{z}|\bs{0} n\rangle\;,
\end{split}
\end{align}
and consequently the Fourier component of the Berry curvature is
\begin{align}
\begin{split}
\lambda_{\bR}&=\int \d \bk\,\e{\i\bk\cdot\bR}F(\bk,M)
=-\i\sum_{n}\langle\bR n|\left(\bR\times\br\right)_{z}|\bs{0} n\rangle
\\
&=-\i\sum_{n}\int (R_{x}r_{y}-R_{y}r_{x})W_{n}(\br-\bR)^{\ast}W_{n}(\br)\d\br\;,
\label{2D_correlation_function}
\end{split}
\end{align}
provided $\bR\neq (0,0)$~\footnote{Note that $\lambda_{\bR}$ at $\bR=(0,0)$ is just the topological invariant, yet the expression in terms of Wannier state becomes troublesome because of the $\bR\times\br$ factor.}.
The Fourier component clearly measures the correlation between occupied Wannier states $|\bR n\rangle$ and $|\bs{0}n\rangle$, and the $\left(\bR\times\br\right)$ factor indicates that it is intimately related to the \enquote{itinerant-circulation} part of the orbital magnetization 
%\notem{this is not used anywhere else, maybe we can remove the $M_\mathrm{IC}$} 
contributed from the edge current but expressed in terms of bulk Wannier states~\cite{Thonhauser05,Ceresoli06,Marzari12}.
The correlation function is gauge-invariant since it is the Fourier transform of the gauge-invariant Berry curvature~\cite{Xiao10}, and a real function since $\lambda_\bR{}^{\ast}=\lambda_{-\bR{}}=\lambda_\bR{}$ in inversion-symmetric systems.
Analogous to Eqs.~\eqref{1D_lambda_equation} and \eqref{lambdar_1D}, the Ornstein--Zernike form of \eq{2D_xi_definition} implies
\begin{align}
\lambda_{\bR}=\lambda(R_{x}/\xi_{x},R_{y}/\xi_{y})\;,
\label{lambdar_2D}
\end{align}
i.e., $\xi_{x}$ and $\xi_{y}$ represent the correlation lengths in the two spatial directions.

\begin{figure}
\centering
\includegraphics[clip=true,width=0.99\columnwidth]{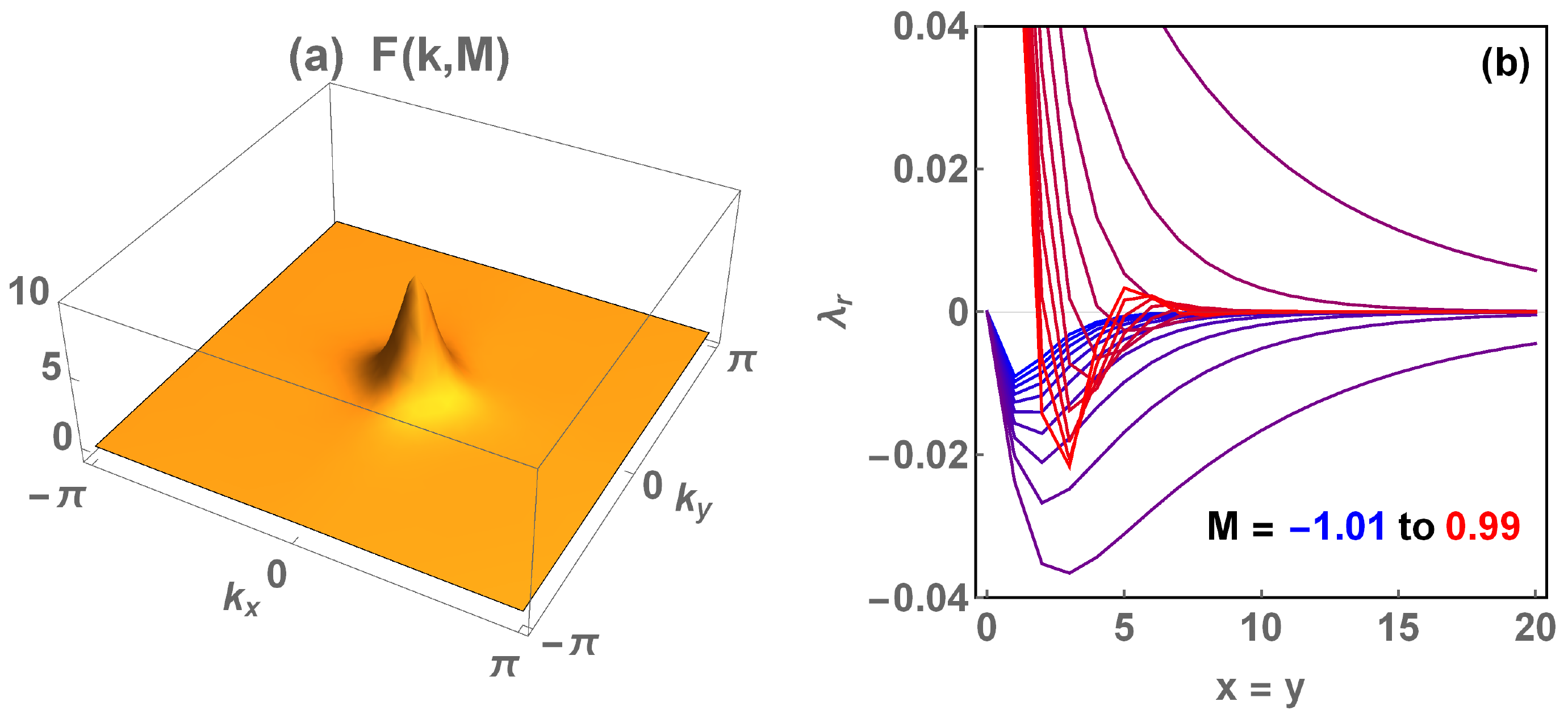}
\caption{(color online) (a) The Curvature function (Berry curvature) $F(\bk,M)$ at $M=0.25$ in the simple model of a 2D Chern insulator described by \eq{2by2_di_simple_Chern}, which diverges at the gap-closing momentum $\bk_{0}=(0,0)$ with a Lorentzian form of \eq{2D_xi_definition} when the system approaches the critical point $M_\rmc=0$.
(b) The Fourier component $\lambda_\br{}$ of $F(\bk,M)$ along the direction $x=y$, which is also the Wannier-state correlation function defined in \eq{2D_correlation_function}.
The correlation function at large distances changes sign between $M\apprle M_\rmc$ and $M\apprge M_\rmc$, and decays with a correlation length that diverges at $M_\rmc$.
The inverse of the correlation length coincides with the width of the peak in (a).} 
\label{fig:2by2_2DTI_Fouriers}
\end{figure}

The correlation function introduced in \eq{lambdar_1D} for 1D and \eq{lambdar_2D} for 2D have the following general trend as the system approaches a topological phase transition. 
Since the Berry connection (at a specific gauge) in 1D and Berry curvature in 2D must diverge at a HSP and change sign at a topological phase transition (such that their integration, the topological invariant, can jump discretely~\cite{Chen16,Chen16_2}), the correlation functions as their Fourier transform must have (i) a diverging correlation length and (ii) a sign change at large distance, as can be seen evidently in Fig.~\ref{fig:SSH_model_lambda_exponents} (b) for the SSH model and Fig.~\ref{fig:2by2_2DTI_Fouriers} (b) in Sec.~\ref{2by2_Dirac} for the 2D Chern insulator. 
Thus, whether a phase transition is topological or not can be identified solely from the correlation function, simply by checking whether (i) and (ii) occur simultaneously or not at the phase transition.

\subsection{Berry curvature of 2D \texorpdfstring{2$\bm{\times}$2}{2x2} Dirac Hamiltonians \label{2by2_Dirac}}

In this and the following subsection, we demonstrate that noninteracting Dirac models in 2D exhibit universal critical exponents regardless the details of the system.
First we consider \xtimesx{2} Dirac Hamiltonians in 2D of the form given in \eq{eq:dirac_1d} with the replacement $k\rightarrow\bk=(k_{x},k_{y})$. The eigenenergies are $E_{\pm}=\pm\sqrt{d_{1}^{2}+d_{2}^{2}+d_{3}^{2}}=\pm d$.
Because we focus here on inversion-symmetric systems, the two degrees of freedom of the Dirac Hamiltonian must have opposite eigenvalues under parity operation (for instance, one represents a $d$ orbital and the other an $f$ orbital, as in the \ac{tki} described in Sec.~\ref{sec:tki}). Choosing the basis that is diagonal under parity operation $P=\sigma_{3}$, the inversion symmetry $PH(\bk)P^{-1}=H(-\bk)$ implies that $d_{1}$ and $d_{2}$ are odd while $d_{3}$ is even in $\bk$, leading to the lowest-order expansion
\begin{subequations}
\label{di_general_form}
\begin{align}
d_{1}&=A_{1x}\delta k_{x}+A_{1y}\delta k_{y}\;,
 \\
d_{2}&=A_{2x}\delta k_{x}+A_{2y}\delta k_{y}\;,
 \\
d_{3}&=\left(M-M_\rmc\right)+B_{x}\delta k_{x}^{2}+B_{y}\delta k_{y}^{2}+B_{xy}\delta k_{x}\delta k_{y}\;,
\end{align}
\end{subequations}
where $\delta\bk=\bk-\bk_{0}$ denotes the distance away from the \ac{hsp}.
The velocity and effective mass parameters $(A_{ix}\commabr A_{iy}\commabr B_{y}\commabr B_{y}\commabr B_{xy})$ depend on the details of the model, and the tuning parameter $M$ again has to enter the parity-even channel since it is a scalar.
The component $d_{0}\sigma_{0}$ is unimportant for the argument below and hence is ignored for simplicity.
The topological invariant of the Dirac Hamiltonian can be calculated from the Berry curvature~\cite{Bernevig13} 
\begin{align}
F(\bk,M)=F_{ij}=\frac{1}{2d^{3}}\epsilon^{abc}d_{a}\partial_{i}d_{b}\partial_{j}d_\rmc\;.
\label{2by2_Berry}
\end{align}
where $\partial_{i}=\partial/\partial \delta k_{i}$.
The expansion of the Berry curvature at $\bk_{0}$ along $i\in\left\{x,y\right\}$ yields \eq{2D_xi_definition} with 
\begin{subequations}
\label{eq:crit_exp_dirac}
\begin{align}
\begin{split}
F(\bk_{0},M)&=\frac{A_{1x}A_{2y}-A_{1y}A_{2x}}{2|M-M_\rmc|\left(M-M_\rmc\right)}
 \\
&\propto\sgn(M-M_\rmc)|M-M_\rmc|^{-2}\;,
\end{split}
 \\
\begin{split}
\xi_{i}&=\left|\frac{8B_{i}\left(M-M_\rmc\right)+3A_{1i}^{2}+3A_{2i}^{2}}{2\left(M-M_\rmc\right)^{2}}\right|^{1/2}
 \\
&\overset{\mathclap{M\to M_\rmc}}{\propto}\quad \left|M-M_\rmc\right|^{-1}\;.
\label{2by2_Dirac_F0_xi}
\end{split}
\end{align}
\end{subequations}
Thus, the critical exponents of the length scale $\xi_{i}\sim\left|M-M_\rmc\right|^{-1}$ and $F(\bk_{0},M)\sim |M-M_\rmc|^{-2}$ are always the same regardless of the microscopic details $(A_{ix},A_{iy},B_{x},B_{y},B_{xy})$ of the model.
One also sees that in general $\xi_{x}\neq \xi_{y}$ if $A_{ix}\neq A_{iy}$ or $B_{x}\neq B_{y}$, yet the scaling law in \eq{scaling_law_2D} is always satisfied.
As the system is driven towards the critical point, similar to the 1D systems demonstrated in Fig.~\ref{fig:xi_versus_Gamma}, there exists a critical region $ (|M-M_\rmc|<|A_{i}^{2}/B_{i}|)$ inside of which the correlation length follows $\xi_{i}\sim|A_{i}/(M-M_\rmc)|$, and outside of which the correlation length is dominated by a different length scale $\xi\sim|B_{i}/(M-M_\rmc)|^{1/2}$, where $A_{i}^{2}=A_{1i}^{2}+A_{2i}^{2}$.
The correspondence between $\xi$ and the edge-state decay length is demonstrated in Appendix~\ref{appendix_xi_decay}, which suggest that the experiments that directly measure $\xi$ by fitting the Berry curvature near the gap-closing momentum, for instance cold-atom experiments~\cite{Jotzu14,Abanin13,Duca15}, can extract the edge-state decay length even in a system with closed boundary condition, and without directly investigating the edge-state wave function.

We proceed to discuss a scaling law that is expected to be satisfied also in interacting systems.
Denoting the critical exponents by%
\begin{subequations}\label{2D_critical_exponent_definition}%
\begin{align}
F(\bk_{0},M)&\propto \sgn(M-M_\rmc)|M-M_\rmc|^{-\gamma}\;,
 \\
\xi_{i}&\propto|M-M_\rmc|^{-\nu_{i}}\;,
\end{align}
\end{subequations}
the argument in \eq{1D_Cprime} generalized to 2D dictates that the curvature function integrated over a small region around $\bk_{0}$ of area $\sim\xi_{x}^{-1}\xi_{y}^{-1}$ converges to a constant:
\begin{align}
\begin{split}
\mathcal{C}_\mathrm{div}&=\int_{-\xi_{x}^{-1}}^{\xi_{x}^{-1}}\frac{\d k_{x}}{2\pi}
\int_{-\xi_{y}^{-1}}^{\xi_{y}^{-1}}\frac{\d k_{y}}{2\pi}
F(\bk,M)
 \\
&=F(\bk_{0},M)\int_{-\xi_{x}^{-1}}^{\xi_{x}^{-1}}\frac{\d k_{x}}{2\pi}\frac{1}{1\pm\xi_{x}^{2}k_{x}^{2}}
\int_{-\xi_{y}^{-1}}^{\xi_{y}^{-1}}\frac{\d k_{y}}{2\pi}\frac{1}{1\pm\xi_{y}^{2}k_{y}^{2}}
 \\
&=\frac{F(\bk_{0},M)}{\xi_{x}\xi_{y}}\times\mathcal{O}(1)=\const \;.
\end{split}
\label{2D_Cprime}
\end{align}   
Thus, the critical exponent of $F(\bk_{0},M)$ is the sum of those of $\xi_{x}$ and $\xi_{y}$
\begin{align}
\gamma=\nu_{x}+\nu_{y}\;,
\label{scaling_law_2D}
\end{align}
which is indeed satisfied by the result for noninteracting systems in \eq{2by2_Dirac_F0_xi}.
This also suggests that as $M\rightarrow M_\rmc$, one can parametrize $F(\bk_{0},M)$ by
\begin{align}
F(\bk_{0},M)\propto\sgn(M-M_\rmc)|M-M_\rmc|^{-\nu_{x}-\nu_{y}}\;.
\end{align}

To see the Wannier-state correlation function, i.e., the Fourier component $\lambda_\bR{}$ of the Berry curvature, we need $\bd(\bk)$ not only near $\bk_{0}$ but in the entire \ac{bz}.
For this purpose, we choose the simple model of a \xtimesx{2} Chern insulator for demonstration, described by~\cite{Bernevig13}%
\begin{subequations}\label{2by2_di_simple_Chern}%
\begin{align}
d_{1}&=\sin k_{x}\;,
 \\
d_{2}&=\sin k_{y}\;,
 \\
d_{3}&=M-2B\left(2-\cos k_{x}-\cos k_{y}\right)\;.
\end{align}
\end{subequations}
The model has three critical points at $M_\rmc\in\{0\commabr 4B\commabr 8B\}$, corresponding to gap-closing at $\bk_{0}\in\{(0,0)\commabr (0,\pi)\commabr (\pi,\pi)\}$.
Focusing on the behavior near the $M_\rmc=0$ critical point, the low-energy sector is described by \eq{di_general_form} with $(A_{1x},A_{2y},B_{x},B_{y})=(1,1,-1,-1)$ after setting the energy unit $B=1$.
According to the calculation in Appendix~\ref{appendix_xi_decay}, the edge state appears when $B_{x}(M-M_\rmc)=-BM=-M<0$; therefore, $M>0$ is the topologically nontrivial state~\cite{Bernevig13}.
The Fourier component of $F(\bk,M)$ calculated from \eq{2by2_Berry} is shown in Fig.~\ref{fig:2by2_2DTI_Fouriers} for the direction $x=y$;
the decaying behavior at large distance is evident, with a correlation length that diverges at the critical point, in agreement with \eq{2by2_Dirac_F0_xi}.
However, the anomalies at short distance---originating from the Fourier transform of $F(\bk,M)$ at momenta far away from $\bk_{0}=(0,0)$---hinder a precise extraction of the correlation length and its critical exponent from this real-space analysis.

\subsection{Universality classes of 2D \texorpdfstring{2$\bm{\times}$2}{2x2} Dirac Hamiltonians \label{Universality_class}}

\begin{figure}
\centering
\includegraphics[width=0.99\linewidth]{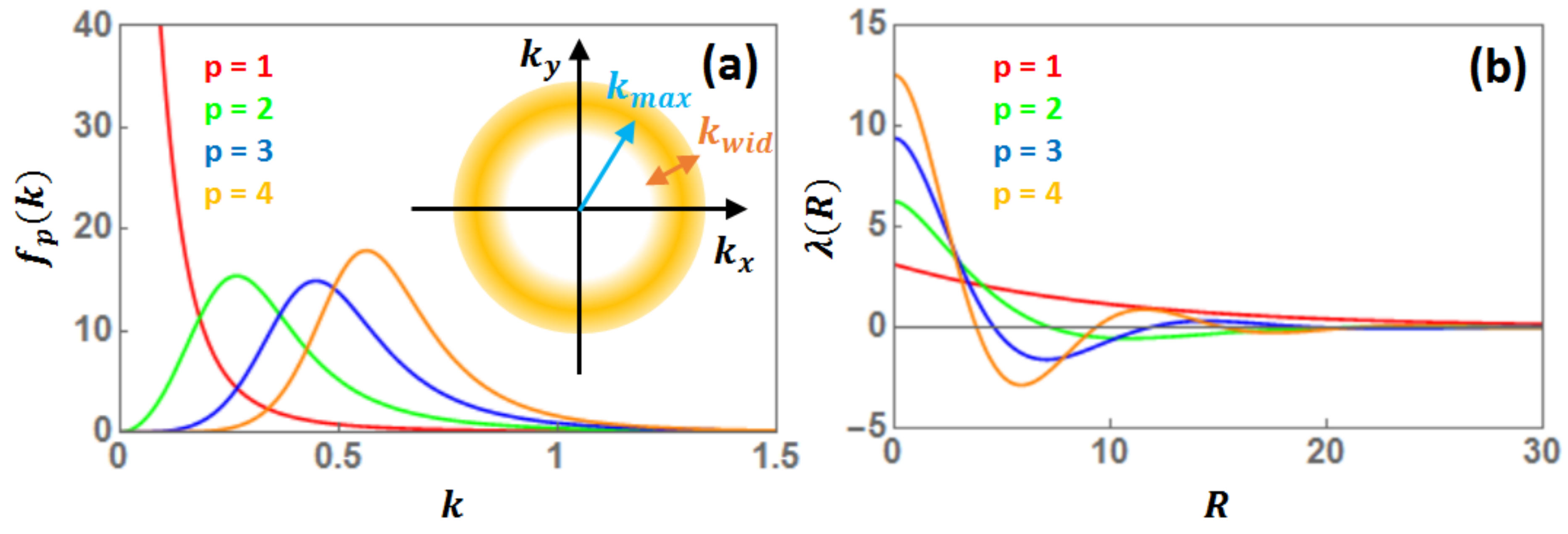}
\caption{ (a) Berry curvature $f_{p}(k)$ and (b) Correlation function $\lambda({\bf R})=\lambda(R)$ of a 2D \xtimesx{2} Dirac model at different orbital symmetries $p\in\{1,2,3,4\}$, at $M=0.1$.
The inset of (a) shows schematically that the maximum of $f_{p}(k)$ at $p\neq 1$ forms a ring surrounding the \ac{hsp}, giving the two momentum scales $k_\mathrm{max}$ and $k_\mathrm{wid}$ in \eq{fmax_kmax_kwid}. 
%\notem{shouldn't $k_\mathrm{wid}$ be the \textbf{half} width at half maximum?}
}
\label{fig:Kr}
\end{figure}

The formalism in Sec.~\ref{2by2_Dirac} can be readily generalized to Dirac models with different orbital symmetries, which have been intensively discussed recently as suitable models for the multi-Weyl semimetals that manifest quadratic, cubic, or higher-order band-touching points~\cite{Xu11,Fang12,Banerjee12,Yang14,Lai15,Jian15,Huang15,DasSarma15,Huang16,Pyatkovskiy16,ChenFiete16,Ahn16}.
Here we demonstrate that the universality class of a 2D \xtimesx{2} Dirac Hamiltonian is determined by its orbital symmetry.
In Sec.~\ref{2by2_Dirac}, we learned that the momentum-dependent corrections to the tuning energy parameters, i.e., the $(B_{x},B_{y},B_{xy})$ terms in \eq{di_general_form}, are unimportant as the system approaches the critical point and are therefore ignored for simplicity.
This leads us to discuss the following \xtimesx{2} Dirac Hamiltonian expressed in polar coordinates, as a low-energy effective model for the topological phase transition that takes place when bulk bands invert at $\bk_{0}$, 
\begin{mysplit}
H(k,\phi)&=\eta_{p}k^p\cos(p\phi)\sigma_x+\eta_{p}k^p\sin(p\phi)\sigma_y+M\sigma_z
 \\
&=\varepsilon_+(k){\hat{\bd}}(k,\phi)\cdot{\boldsymbol\sigma}\;,
\label{eq:2by2_Dirac_general_p}
\end{mysplit}
where momentum $\bk$ is the distance from $\bk_{0}$.
The positive integer $p\in\{1,2,3,\dots\}$ represents the orbital symmetry $\left\{p,d,f,\dots\right\}$ of the Hamiltonian and $\eta_{p}>0$ is a positive quantity that keeps track of the dimension.
The formalism in Sec.~\ref{2by2_Dirac} with parameters $A_{1y}=A_{2x}=B_{x}=B_{y}=B_{x,y}=0$ is equivalent to $p=1$ ($p$-wave) in this section, where $\eta_{p}=A_{1x}=A_{2y}=v_\mathrm{F}$ is the Fermi velocity.
The energy dispersion for a general $p$ is given by
\begin{equation}
\varepsilon_{\pm}(k)=\pm\sqrt{M^2+\eta_{p}^2k^{2p}} \;.
\end{equation}
The ${\hat{\bd}}$ vector in \eq{eq:2by2_Dirac_general_p} together with \eq{2by2_Berry} yields the Berry curvature 
\begin{mysplit}
f_{p}(k_x,k_y)&={\hat{\bd}}\cdot\left(\frac{\partial{\hat{\bd}}}{\partial k_x}\times\frac{\partial{\hat{\bd}}}{\partial k_y}\right)
 \\
&=\frac{k^{-2 + 2 p} M p^2\eta_{p}^{2}}{(\eta_{p}^{2}k^{2 p} + M^2)^{3/2}}=f_{p}(k)\;,
\label{Berry_curvature_at_p}
\end{mysplit}
which does not depend on the polar angle $\phi$ as shown in Fig.~\ref{fig:Kr} (a).
The topological invariant is given by
\begin{equation}
{\cal C}=\frac{1}{4\pi}\int_0^{2\pi}\d\phi\int_0^{\infty}k\d k\;f_p(k)=\frac{p}{2}\sgn(M)\;,
\label{deltaC_general}
\end{equation}
normalized in such a way that the change of the topological invariant across the topological phase transition is $\Delta{\cal C}={\cal C}(M>0)-{\cal C}(M<0)=p$.
This normalization is in accordance with the low-energy description of 2D Chern insulators, in which the topological invariant changes by $\Delta{\cal C}=p=1$ at a topological phase transition when bulk bands invert at one \ac{hsp}~\cite{Bernevig13}.

%Equation~\eqref{deltaC_general} demonstrates the correspondence $\Delta{\cal C}=p$ between the orbital symmetry and the change of topological invariant across a topological phase transition, provided the bulk band is inverted at one \ac{hsp}. \notem{This seems to be an unnecessary repetition of the previous sentence}
In certain models, the bulk bands can be inverted simultaneously at multiple \acp{hsp} when going through a topological phase transition.
In such a case, the total change of the topological invariant is given by the sum of the $\Delta{\cal C}$ at each HSP. %\notem{here we may need to be careful about the signs: it is possible that one \ac{hsp} contributes $+p$ while another one $-p$, leading to a total change of 0 instead of $2p$}
For instance, in 2D Chern insulators, the bulk band can be inverted simultaneously at $\bk_{0}=(0,\pi)$ and $\bk_{0}=(\pi,0)$ at a particular critical point, and hence the topological invariant of the whole system changes by 2 since each \ac{hsp} contributes 1~\cite{Bernevig13}.

Systems with different orbital symmetries $p$ belong to different universality classes, as can be seen by comparing the critical exponents at different $p$.
For the sake of introducing the critical exponents, we consider our continuous Hamiltonian, \eq{eq:2by2_Dirac_general_p}, as a low-energy effective model of a 2D lattice model in which the Wannier states $|{\bf R}\rangle$ can be introduced via \eq{Wannier_basis} (for simplicity, we drop the band index $n$ by assuming only one filled band), and consequently the correlation function in \eq{2D_correlation_function} can be introduced as
\begin{mysplit}
\lambda({\bf R})&=-\i\langle{\bf R}|({\bf R}\times{\bf r})_{z}|{\bf 0}\rangle
 \\
&=\frac{1}{(2\pi)^2}\int \d k_x \d k_y \,\e{\i\bk\cdot{\bf R}}f_p(k_{x},k_{y})
 \\
&\approx\frac{1}{2\pi}\int_0^{\infty}\d k\, k J_0(R k) f_p(k)\;,
\label{correlation_function}
\end{mysplit}
where $J_0(x)$ is the 0\textsuperscript{th}-order Bessel function.
In the last line of \eq{correlation_function}, we approximate the 2D integral within the \ac{bz} by an integral over the infinite polar plane, and use the Berry curvature in \eq{Berry_curvature_at_p}.
This approximation yields a closed expression of $\lambda({\bf R})$ for a given $p$, which is a function of $R/\xi_p$ with the correlation length given by
\begin{equation}
\xi_p=\left(\frac{\eta_{p}}{|M|}\right)^{1/p}\;,
\label{xi_scaling}
\end{equation}
indicating that the critical exponent for the correlation length is $1/p$.
Figure~\ref{fig:Kr} shows the correlation function at several values of $p$.

The Berry curvature $f_{p}(k)$ at $p=1$ has a maximum at $k=0$, or equivalently a maximum at the \ac{hsp} $\bk_{0}$ as shown in Fig.~\ref{fig:2by2_2DTI_Fouriers} (a), hence the scaling law deduced in \eq{scaling_law_2D} is expected to be satisfied.
At $p\neq 1$, however, the maximum of $f_{p}(k)$ occurs at finite $k$, meaning that the maximum of $f_{p}(k)$ forms a ring around the \ac{hsp} in momentum space.
This motivates us to introduce the following scaling law for $p\neq 1$.
We denote the momentum at which the maximum of the Berry curvature $f_{p,\mathrm{max}}(k_\mathrm{max})$ occurs as $k_\mathrm{max}$ (radius of the ring), and the half width at half maximum of the Berry curvature as $k_\mathrm{wid}$ (width of the ring), as shown schematically in Fig.~\ref{fig:Kr} (a).
Denoting critical exponents of $f_{p,\mathrm{max}}$ and the corresponding length scales by%
\begin{subequations}
\begin{align}
f_{p,\mathrm{max}}(k_\mathrm{max})&\sim M^{-\gamma}\;,
 \\
\xi_\mathrm{max}\sim k_\mathrm{max}^{-1} &\sim |M|^{-\nu_{0}}\;,
 \\
\xi_\mathrm{wid}\sim k_\mathrm{wid}^{-1} &\sim |M|^{-\nu_{1}}\;,
\end{align}\label{fmax_kmax_kwid}%
\end{subequations}
we observe that when the system is approaching the critical point, the integral of the Berry curvature remains unchanged since it is the topological invariant.
Therefore, the maximum times the area of the ring should remain the same
\begin{mysplit}
{\cal C}_\mathrm{div}&\sim f_{p,\mathrm{max}}\left[\pi\left(k_\mathrm{max}+\frac{k_\mathrm{wid}}{2}\right)^{2}-\pi\left(k_\mathrm{max}-\frac{k_\mathrm{wid}}{2}\right)^{2}\right]
 \\
&\sim \frac{2\pi f_{p,\mathrm{max}}}{\xi_\mathrm{max}\xi_\mathrm{wid}}
=\const\;,
\end{mysplit}
and consequently the scaling law
\begin{mysplit}
\gamma=\nu_{0}+\nu_{1}
\label{scaling_law_higher_p}
\end{mysplit}
should be satisfied.
The Berry curvature in \eq{Berry_curvature_at_p} gives $\gamma=2/p$ and $\nu_{0}=\nu_{1}=1/p$, which indeed satisfy the scaling law.
The critical exponents $\nu_{0}$ and $\nu_{1}$ are in accordance with that of the correlation length in \eq{xi_scaling}, indicating that there is essentially only one length scale in the problem.
The critical behaviors of various quantities are summarized in Table \ref{tab:universality_class}.

\begin{table}[ht]
  \centering
  \caption{Summary of the critical behavior of various quantities for a 2D \xtimesx{2} Dirac model with orbital symmetry $p$, where $i\in\{x,y,0,1\}$ labels the correlation lengths defined in Eqs.~\eqref{2D_critical_exponent_definition} and \eqref{fmax_kmax_kwid}.}
  \label{tab:universality_class}
\begin{ruledtabular}
  \begin{tabular}{ll}
Dispersion near \ac{hsp} & $\varepsilon_{\pm}(k)=\pm\left(M^2+\eta_{p}^2k^{2p}\right)^{1/2}$ \\
Berry curvature maximum & $f_{p,\mathrm{max}}\sim|M-M_{c}|^{-\gamma}$,$\gamma=\frac{2}{p}$ \\
Correlation length & $\xi_{i}\sim|M-M_{c}|^{-\nu_{i}}$,\;$\nu_{i}=\frac{1}{p}$ \\
Scaling law for $p=1$ & $\gamma=\nu_{x}+\nu_{y}$ \\
Scaling law for $p\neq 1$ & $\gamma=\nu_{0}+\nu_{1}$ \\
Change of invariant & $\Delta{\cal C}=p$
  \end{tabular}
\end{ruledtabular}
\end{table}

For $p\neq 1$, the fact that the Berry curvature $f_{p}(k)$ has an extremum at finite $k$, poses a challenge to a previously proposed \ac{rg} scheme~\cite{Chen16}, as it relies on the fact that the scaling function, which is previously identified as the Berry curvature, has an extremum at $k=0$.
This is true for all the models examined previously whose low-energy sectors are Dirac models with $p=1$~\cite{Chen16,Chen16_2}.
For a general $p$, we propose to use the $(2p-2)$\textsuperscript{th} derivative of the Berry curvature as the scaling function
\begin{mysplit}
F(\bk,M)=\partial_{k}^{2p-2}f_{p}(k)
\label{scaling_function_general_p}
\end{mysplit}
in the \ac{rg} procedure
\begin{mysplit}
F(\delta\bk,M)=F({\bf 0},M^{\prime})\;, 
\label{RG_procedure}
\end{mysplit}
where $\delta\bk$ is a small deviation away from the \ac{hsp} (the \ac{hsp} is implicitly taken as the origin $\bk_{0}={\bf 0}=(0,0)$).
Solving for the new $M^{\prime}$ at a given $M$ iteratively yields the \ac{rg} flow that distinguishes the topological phase transitions.
The choice of the scaling function given in \eq{scaling_function_general_p} is justified because it satisfies the following criteria for the \ac{rg} procedure: (1) The scaling function has an extremum at $k=0$, i.e., $\partial_{k}F(\bk,M)|_{k=0}=\partial_{k}^{2p-1}f_{p}(k)|_{k=0}=0$.
(2)~The scaling function diverges at the critical point $M_{c}=0$ and has opposite sign as $M\rightarrow M_{c}^{+}$ and $M\rightarrow M_{c}^{-}$.
(3)~Its integral is a topological invariant 
\begin{mysplit}
{\cal C}^{\prime}&=\frac{1}{4\pi}\int_0^{2\pi}\d\phi\int_0^{\infty}k\d kF(\bk,M)
 \\
&=
\begin{cases}
\frac{1}{2}\sgn(M)={\cal C} & \text{for }p=1\;, \\
0 & \text{for }p\neq 1\;.
\end{cases}
\end{mysplit}
With these criteria satisfied, the \ac{rg} procedure depicted in \eq{RG_procedure} can be used to judge topological phase transitions because it makes $F(\bk,M)$ flow away from the critical-point configuration and converge to the fixed-point configuration $F_\mathrm{f}(\bk,M)$ without changing the topological invariant ${\cal C}^{\prime}$ in the whole procedure~\cite{Chen16}.

Since the Berry curvature of our low-energy effective theory, \eq{Berry_curvature_at_p}, does not depend on the polar angle $\phi$, the scaling direction can be taken along any radial direction $\delta{\hat\bk}={\hat\bk}$.
We can substitute the Berry curvature given in \eq{Berry_curvature_at_p} and the scaling function of \eq{scaling_function_general_p} into \eq{RG_procedure} and expand the equation for a small $\delta k=|\delta\bk|$ and $\delta M=M^{\prime}-M$ to the lowest nonvanishing order.
This procedure leads to the generic \ac{rg} equations
\begin{mysplit}
\frac{\d M}{\d l}\equiv\frac{\delta M}{\delta k^{2p}}=
\frac{1}{(2p)!}\frac{\partial_{k}^{2p}F(\bk,M)|_{\bk={\bf 0}}}{\partial_{M}F({\bf 0},M)}=\frac{A_{p}}{M}\;,
\label{generic_RG_equation}\end{mysplit}
where $A_p$ is a $p$-dependent numerical factor---$(A_1\commabr A_2\commabr A_3\commabr A_4\commabr \dots)=(3/4\commabr 45/4\commabr 315/2\commabr 9009/4\commabr \dots)$---and the scaling parameter is $\d l\equiv \delta k^{2p}$ since it is the lowest nonvanishing order in the expansion.
From \eq{generic_RG_equation}, we see that the critical point lies at $M_{c}=0$ for all $p$ as expected, since the Dirac Hamiltonian we start with, \eq{eq:2by2_Dirac_general_p}, has a band inversion at $M_{c}=0$.
The lowest-order expansion of the scaling function in $\delta k$ is
\begin{mysplit}
F(\delta\bk,M)=\frac{x_{p}}{M|M|}-\frac{y_{p}}{M^{3}|M|}\delta k^{2p}+\dots
\approx\frac{F({\bf 0},M)}{1+\frac{y_{p}\delta k^{2p}}{x_{p}M^{2}}}\;,
\label{scaling_fn_expansion}
 \\
\end{mysplit}
which takes the Lorentzian form with a half width at half maximum that scales like $\delta k_{H}\sim |M|^{1/p}$, where $x_{p}$ and $y_{p}$ are $p$-dependent numerical factors.
Consequently, the length scale $\xi\sim\delta k_{H}^{-1}\sim |M|^{-1/p}$ extracted from the scaling function in \eq{scaling_fn_expansion} coincides with the correlation length in \eq{fmax_kmax_kwid}, meaning that there is still only one length scale in the problem whose critical behavior is $|M|^{-1/p}$.

\section{Critical exponents in mirror planes of a correlated topological insulator}\label{sec:tki}

We further predict that the scaling laws proposed in \eq{1D_critical_exponent_summary} for 1D and Eqs.~\eqref{scaling_law_2D} and \eqref{scaling_law_higher_p} for 2D have to be satisfied even when the topological phase transitions are driven by interactions, as demonstrated in the following example at the mean-field level.
The model we consider is a simplified model for a \ac{tki}, such as samarium hexaboride, on a simple cubic lattice~\cite{Dzero10, Legner14}:
\begin{subequations}
\begin{align}
H&=H_0+H_\mathrm{hyb}+H_\mathrm{int}\;,\\
H_0&=
\begin{aligned}[t]
&\sum_{i}\ef\,f^\dag_if^{\nodag}_i-\sum_{\Nn{i,j}}\left(\td\,c^\dag_ic^{\nodag}_j+\tf\,f^\dag_if^{\nodag}_j+\hc\right)\\
&-\sum_{\Nnn{i,j}}\left(\tdd\,c^\dag_ic^{\nodag}_j+\tff\,f^\dag_if^{\nodag}_j+\hc\right)\;,
\end{aligned}\\
H_\mathrm{hyb}&=\sum_{\alpha=x,y,z}\sum_{\Nn{i,j}_\alpha}\!\!\left[\i V c_i^\dag\sigma_\alpha f^{\nodag}_j+\i V f_i^\dag\sigma_\alpha c^{\nodag}_j+\hc\right],\\
H_\mathrm{int}&=\sum_iU\,f^\dag_{i\uparrow}f^{\nodag}_{i\uparrow}\,f^\dag_{i\downarrow}f^{\nodag}_{i\downarrow}\;.\label{eq:Hint}
\end{align}\label{eq:model_tki}%
\end{subequations}
Here, $c^\dag$, $c$ and $f^\dag$, $f$ are creation- and annihilation-operators for $d$ and $f$ electrons, respectively; 
$H_0$ represents the onsite energy of $f$ electrons as well as hopping of $d$ and $f$ electrons between nearest neighbors ($\Nn{i,j}$) and next-to-nearest neighbors ($\Nnn{i,j}$); 
$H_\mathrm{hyb}$ represents the (odd-parity) hybridization between $d$ and $f$ orbitals; 
and $H_\mathrm{int}$ is the onsite repulsion for $f$ orbitals.

It was shown in Ref.~\onlinecite{Legner14} that a mean-field treatment based on the Kotliar--Ruckenstein slave-boson scheme~\cite{Kotliar86} can transform this Hamiltonian into a non-interacting Hamiltonian with renormalized parameters, 
% \begin{subequations}
	\begin{align}
		\tf\to z^2\tf\;,\quad
		\tff\to z^2\tff\;,\quad
		V\to zV\;,\quad
		\ef\to\ef+\lambda\;.
	\end{align}
% \end{subequations}
When tuning the onsite-repulsion $U$, the gap can close at the \acp{hsp} $\bf \Gamma$, \textbf{X}, \textbf{M}, or \textbf{R}, and one can therefore observe topological phase transitions.

While the system is three-dimensional, the mirror symmetries of the simple cubic lattice allow the definition of three distinct \acp{mcn}~\cite{Teo08}:
In the three mirror-invariant planes in momentum space, $k_z=0$, $k_z=\pi$, and $k_x=k_y$~\footnote{All other mirror-invariant planes are related to those three planes by symmetry operations.}, the energy eigenstates may be chosen to be simultaneous eigenstates of the respective mirror operator, $M |u_a^{\pm}(\tilde \bk)\rangle=\pm\i|u_a^{\pm}(\tilde \bk)\rangle$, where the momentum $\tilde\bk$ lies in this mirror-invariant plane.
Considering only the states $|u^+_a(\tilde\bk)\rangle$ for the Berry curvature, we can then calculate the three \acp{mcn} $C_{k_z=0}^+$, $C_{k_z=\pi}^+$, and $C_{k_x=k_y}^+$~\cite{Legner14}.
Close to the \ac{hsp} where the gap closes, we can map the states with mirror eigenvalue $+\i$ in the different mirror planes to two-dimensional Dirac models, where the mass parameter $M$ is a function of the model parameters.

\begin{figure}
	\centering
	\includegraphics{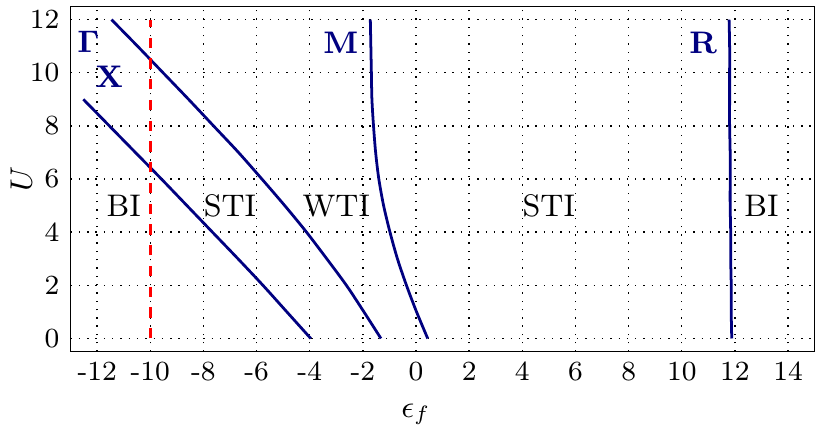}
	\caption{(color online) Phase diagram of the simple model for a \ac{tki} as a function of $\ef$ and $U$ for $\td=1$, $\tdd=-0.4$, $\tf=-0.1$, $\tff=0.04$, and $V=0.4$ from a mean-field treatment with Kotliar--Ruckenstein slave bosons~\cite{Legner14} showing strong-topological-insulator (STI), weak-topological-insulator (WTI), and (trivial) band-insulator phases (BI).	
Phase transitions occur whenever the gap closes at one of the \acp{hsp} $\bf \Gamma$, \textbf{X}, \textbf{M}, or \textbf{R}.
The dashed red line shows the parameters of the plots in Fig.~\ref{fig:tki_crit-exp}.}
	\label{fig:phases_tki}
\end{figure}

\begin{figure}
	\centering
	\includegraphics{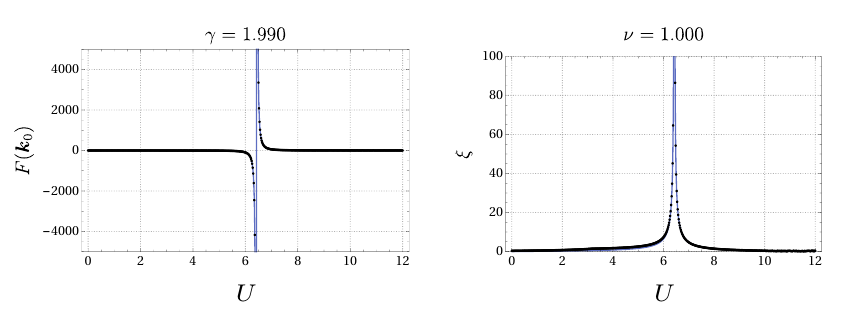}
	
	\caption{$F(\bk_0)$ and $\xi$ as a function of $U$ with fit for $\bk_0=\mathbf{X}$ in the plane $k_z=0$, $\ef=-10$, and other parameters as in Fig.~\ref{fig:phases_tki}.
The parameters correspond to the dashed red line in Fig~\ref{fig:phases_tki}.
The phase transition at $\bf\Gamma$ is not visible, as we are only considering $\bk_0=\bf X$.}
	\label{fig:tki_crit-exp}
\end{figure}

Keeping all other model parameters fixed, we can consider the phase diagram as a function of $\ef$ and $U$, as shown in Fig.~\ref{fig:phases_tki}~\cite{Legner14}.
In this setup, we can observe topological phase transitions by varying either $\ef$ or $U$ and fit the critical exponents $\gamma$ and $\nu$ for different \acp{hsp} (see Figs.~\ref{fig:phases_tki} and \ref{fig:tki_crit-exp}).
Close to the transition, we can approximate the system by a \xtimesx{2} Dirac model where the mass term is a function of either $\ef$ or $U$.
In all cases, we find that $M-M_\rmc$ is approximately proportional to $\ef-{\ef}_\rmc$ or $U-U_\rmc$ such that the critical exponents $\gamma\approx 2$, $\nu_{x}\approx\nu_{y}\approx 1$ are close to the values of the non-interacting case with orbital symmetry $p=1$ (see Table \ref{tab:universality_class}).
In particular, the scaling law~\eqref{scaling_law_2D} is always satisfied up to numerical precision.

\begin{figure}[tb]
	\centering
	\includegraphics{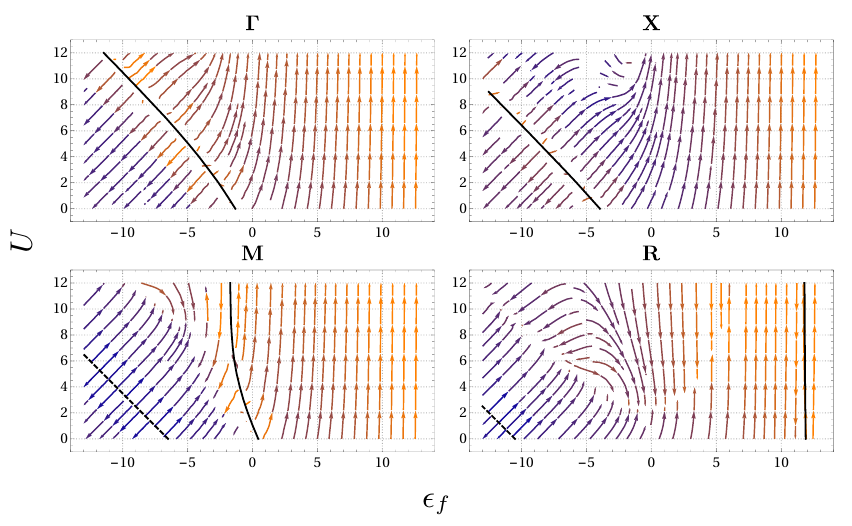}

	\caption{(color online) Plots of the \ac{rg} flow for the same parameters as in Fig.~\ref{fig:phases_tki} for the four \acp{hsp} $\bf \Gamma$, \textbf{X}, \textbf{M}, and \textbf{R}.
The color indicates the flow rate, where blue denotes small and orange high flow rates.
	The thick black line is the critical line where the gap closes and is the same as those in Fig.~\ref{fig:phases_tki}.
	Unstable fixed points are shown by dashed lines for $\bk_0\in\{\mathbf{M},\mathbf{R}\}$}.
	\label{fig:RG_tki}
\end{figure}

The \ac{rg} flow in the parameter space $\bs M=(M_1,M_2)=(\ef ,U)$ can be obtained by~\cite{Chen16} 
\begin{subequations}
	\begin{align}
		F(\bk_{0},\epsilon_{f}^{\prime},U)&=F(\bk_{0}+\delta\bk,\epsilon_{f},U)\;,
		\\
		F(\bk_{0},\epsilon_{f}',U^{\prime})&=F(\bk_{0}+\delta\bk,\epsilon_{f}',U)\;.
	\end{align}
\end{subequations}
Since the model has linear band crossing ($p=1$ in Table~\ref{tab:universality_class}), \eq{generic_RG_equation} yields
\begin{equation}
	\frac{M_i'-M_i}{\delta k^2}=\frac{\left.\partial^2_{k}F(\bk,\bs M)\right|_{\bk=\bk_0}}{2\partial_{M_i} F(\bk_0,\bs M)}
\end{equation}
for $i=1,2$, where the first derivative in $M_{i}$ and the second derivative at $\bk$ can be obtained numerically in a very efficient manner by computing only three points $F(\bk_{0},M_{i})$, $F(\bk_{0}+\delta\bk,M_{i})$, and $F(\bk_{0},M_{i}+\delta M_{i})$.
%This demonstrates the advantage of this \ac{rg} scheme compared to brute force integration of \eq{1D_winding_number}, which would require to calculate $F(\bk,M_{i})$ at all $\bk$'s in the \ac{bz}.
%\notem{I don't really agree with these two sentences; the calculation of a (mirror) Chern number is computationally extremely fast; the problem with this model is the free-energy minimization which cannot be avoided by using the RG scheme; if we only want to detect phase transitions, we can simply check the band gap at the \acp{hsp}}
The two-dimensional \ac{rg} flow is then deduced from the vector field
\begin{equation}
	\bs V(\ef,U)=
	\begin{pmatrix}
		\ef '-\ef \\
		U'-U
	\end{pmatrix}
\end{equation}
and is plotted in Fig.~\ref{fig:RG_tki} for $\bk_0\in\{\mathbf{\Gamma}, \mathbf{X}, \mathbf{M}, \mathbf{R}\}$ and $\delta k=\frac{\pi}{100}$.
The phase transitions can be identified as the critical lines of the \ac{rg} flow, i.e., sources of the \ac{rg} flow where $\norm{\bs{V}}\to\infty$, and are shown as black lines.
This model also displays \emph{unstable fixed points}, sources of the \ac{rg} flow but with $\norm{\bs{V}}\to 0$, where no topological phase transition takes place.
Examples are shown in the plots for \textbf{M} and \textbf{R} as dashed lines.
Note also that $(U'-U) \to \infty$ for large $\ef$ as there $\partial_U F(\bk_0)\to0$.

\section{Conclusions}\label{sec:concl}

In summary, we reveal the following intriguing aspects of topological phase transitions in \xtimesx{2} Dirac Hamiltonians.
The first point addresses the correlation function and correlation length.
We suggest that the Fourier component $\lambda_\bR{}$ of the Berry connection in 1D and that of the Berry curvature in 2D represent correlation functions.
The correlation function in 1D measures the charge-polarization correlation between filled-band Wannier states at different positions, whereas in 2D it measures a correlation between filled-band Wannier states that is intimately related to the itinerant-circulation part of orbital magnetization.
These Wannier-state correlation functions at large distance decay with a correlation length, which can be measured in 2D since it is gauge invariant in this case.
For instance, it could be measured in the \acp{ti} that are artificially engineered in 2D optical lattices.
In the topologically nontrivial state, the correlation length coincides with the decay length of the edge state up to a constant prefactor.
%{\color{green}Does this remain true also for $p>1$? Naively, I expect the decay length of the edge state to scale with the inverse energy gap.}
Despite this coincidence, the correlation length seems to be more general for the characterization of topological phases than the decay length of the edge state, since it manifests even in the topologically trivial state and in periodic boundary conditions. In addition, the correlation function can be used to identify whether a phase transition is topological or not, since the correlation function must have a diverging correlation length and a sign change at large distance at a phase transition that is topological.

 The second feature are the critical exponents and the scaling laws that constrain them.
The noninteracting Dirac Hamiltonians with only one even-momentum channel are shown to have universal critical exponents regardless of the microscopic details of the Hamiltonian.
In essence, this is because the tuning energy parameter can only enter the channel that is even around the gap-closing momentum, rendering a curvature function that has a particular power-law dependence on the tuning energy parameter. 
%regardless of the nature of the tuning energy parameter. \notem{some repetition of \enquote{tuning energy parameter}}
On the other hand, the saturation of the integral of the curvature function near the gap-closing momentum leads to scaling laws, which may be verified experimentally in the gauge-invariant cases in 2D.
Using a model of a \ac{tki}, we demonstrate that the scaling laws should be satisfied even in interacting models, and irrespective of whether the tuning energy parameter $M$ is a interacting or noninteracting term.
Finally, the orbital symmetry of 2D \xtimesx{2} Dirac models fixes the critical exponents of various quantities, as summarized in Table \ref{tab:universality_class}, allowing the notion of universality class to be introduced.

\section{Acknowledgments}
The authors acknowledge fruitful discussions with A.~P.~Schnyder, S.~Huber, G.~Jotzu, P.~Jakubczyk, S.~Ryu, and A.~V.~Balatsky.
This work has been financially supported by a grant of the Swiss National Science Foundation.

\appendix

\section{Correspondence between correlation length and edge-state decay length \label{appendix_xi_decay}}

Here, we elaborate the correspondence between the correlation length $\xi$ and the decay length of the edge state in the topologically nontrivial state, using the well-established analysis of Dirac Hamiltonians projected to real space~\cite{Konig08,Zhou08,Linder09,Lu10,Shen12,Qi11}.
First, let us consider the 1D \xtimesx{2} Dirac Hamiltonian discussed in Sec.~\ref{1D_2by2_section}.
For a model described by  the gauge choice of \eq{1D_2by2_wave_function} and defined in the $x>0$ half-space,
we aim at calculating the zero-energy edge state whose wave function is localized near $x=0$ and satisfies, after replacing $\delta k=-\i \partial_{x}$, 
\begin{align}
(\bd\cdot\bsigma)\psi=\left\{-\i A\sigma_{y}\partial_{x}
+\left[(M-M_\rmc)-B\partial_{x}^{2}\right]\sigma_{x}\right\}\psi=0\;.
\label{1D_2by2_edge_Dirac}
\end{align}
After multiplying this equation by $\sigma_{y}$, we see that the wave function $\psi=\chi_{\eta}\phi(x)$ is an eigenstate of $\sigma_{z}$, i.e., $\sigma_{z}\chi_{\eta}=\eta\chi_{\eta}$ with $\eta=\pm 1$.
Using the ansatz for the spatial dependence $\phi(x)\propto\e{-x/\xi}$\;, we obtain the solutions
\begin{align}
\xi_{\pm}^{-1}=\frac{1}{2}\left[-\eta\frac{A}{B}\pm\sqrt{\left(\frac{A}{B}\right)^{2}
+\frac{4(M-M_\rmc)}{B}}\;\right]\;.
\label{1D_2by2_edge_xi}
\end{align}
The requirement of a positive decay length $\xi_{+}^{-1}+\xi_{-}^{-1}=-\eta A/B>0$ demands $\eta=-\sgn(A/B)$, i.e., the eigenvalue of $\chi_{\eta}$ depends on the sign of $A/B$. The longer of the two, $\xi_{-}$, represents the decay length of the edge state which, when $M\rightarrow M_\rmc$, approaches
\begin{align}
\xi_{-}=-\sgn(B(M-M_\rmc))\left|\frac{A}{M-M_\rmc}\right|\approx\xi\;.
\label{1D_2by2_edge_correlation_correspondence}
\end{align}
One sees that the edge state only appears in the topologically nontrivial state~\cite{Qi11} $B(M-M_\rmc)<0$, and has a decay length $\xi_{-}>0$ that coincides with the correlation length in \eq{1D_2by2_A0_xi} when $M\rightarrow M_\rmc$.

One may further consider the edge state at the interface between a topologically trivial and a nontrivial state, i.e., $M-M_{c}$ changes sign at the interface.
This can be easily obtained from the analysis in Eqs.~\eqref{1D_2by2_edge_Dirac} to \eqref{1D_2by2_edge_correlation_correspondence} by considering a 1D Dirac model in the real space $-\infty<x<\infty$ that has a position dependent mass term $M(x)-M_{c}=M_{1}\theta(x)+M_{2}\theta(-x)$, where $\theta(x)$ is the step function.
$M_{1}$ and $M_{2}$ must have opposite sign, as we shall see below.
Using the ansatz for the zero-energy edge state~\cite{Shen12} 
\begin{mysplit}
\psi\propto\chi_{\eta}\exp\left(-\eta\int_{0}^{x}\frac{dx^{\prime}}{\xi(x^{\prime})}\right)\;,
\label{interface_wave_function}
\end{mysplit}
where $\sigma_{z}\chi_{\eta}=\eta\chi_{\eta}=\pm\chi_{\eta}$, the same analysis leads to \eq{1D_2by2_edge_xi} with the replacement $-\eta A/B\rightarrow -A/B$.
For $-A/B>0$, the longer one of the two solutions is
\begin{mysplit}
\xi_{-}(x)&=-\sgn(B(M(x)-M_\rmc))\left|\frac{A}{M(x)-M_\rmc}\right|\;,
 \\
\eta&=
\begin{cases}
-\sgn(B(M(x)-M_\rmc)) & \mathrm{for}\; x>0 \\
\sgn(B(M(x)-M_\rmc)) & \mathrm{for}\; x<0
\end{cases}
\label{interface_xi_case1}
\end{mysplit}
The sign of $\eta$ is chosen such that the multiple $\eta/\xi_{-}$ makes the wave function in \eq{interface_wave_function} vanish at $x=\pm\infty$.
Likewise, for $-A/B<0$, the longer one is
\begin{mysplit}
\xi_{+}(x)&=\sgn(B(M(x)-M_\rmc))\left|\frac{A}{M(x)-M_\rmc}\right|\;,
 \\
\eta&=
\begin{cases}
\sgn(B(M(x)-M_\rmc)) & \mathrm{for}\; x>0 \\
-\sgn(B(M(x)-M_\rmc)) & \mathrm{for}\; x<0
\end{cases}
\label{interface_xi_case2}
\end{mysplit}
For either case, one sees that $M(x)-M_{c}$ must be of opposite sign between $x>0$ and $x<0$ to have the same value of $\eta$, i.e., the edge state is in the same eigenstate of $\sigma_{z}$ on both sides.
Moreover, from Eqs.~\eqref{interface_xi_case1} and \eqref{interface_xi_case2} one sees that the edge-state decay length for both the topologically nontrivial side ($B(M(x)-M_{c})<0$) or the trivial side ($B(M(x)-M_{c})>0$) coincides with the correlation length in \eq{1D_2by2_A0_xi}.

For the 2D \xtimesx{2} Dirac Hamiltonian discussed in Sec.~\ref{2by2_Dirac}, the procedure is similar.
Consider the model defined in the half-space $x>0$ of the 2D plane.
We look for the solution of the edge state at transverse momentum $\delta k_{y}=0$ and zero energy $E=0$.
Using $\delta k_{x}=-\i \partial_{x}$ leads to%
\begin{align}%
\begin{multlined}%
\big\{-\i \left(A_{1x}\sigma_{x}+A_{2x}\sigma_{y}\right)\partial_{x}\\
{}+\left[(M-M_\rmc)-B_{x}\partial_{x}^{2}\right]\sigma_{z}\big\}\psi=0\;.
\label{2by2_edge_differential_eq}
\end{multlined}
\end{align}
Performing a rotation of the $\left\{\sigma_{x},\sigma_{y}\right\}$ components by%
\begin{subequations}%
\begin{align}
\begin{pmatrix}
\sigma_{x} \\
\sigma_{y}
\end{pmatrix}&=
\begin{pmatrix}
\cos\theta & \sin\theta \\
-\sin\theta & \cos\theta
\end{pmatrix}
\begin{pmatrix}
\sigma_{x}^{\prime} \\
\sigma_{y}^{\prime}
\end{pmatrix}
\;,
 \\
\theta&=\arctan\left(\frac{-A_{2x}}{A_{1x}}\right)\;,
\end{align}
\end{subequations}
leads to 
\begin{align}
\left\{-\i A_{x}\sigma_{x}^{\prime}\partial_{x}
+\left[(M-M_\rmc)-B_{x}\partial_{x}^{2}\right]\sigma_{z}^{\prime}\right\}\psi=0\;,
\label{2by2_edge_differential_eq_rotated}
\end{align}
with $A_{x}=A_{1x}\cos\theta-A_{2x}\sin\theta$.
Multiplying the equation by $\sigma_{x}^{\prime}$, we see that the wave function is an eigenstate of $\sigma_{y}^{\prime}$.
The ansatz $\psi\propto\chi_{\eta}\e{-x/\xi}$, with $\sigma_{y}^{\prime}\chi_{\eta}=\eta\chi_{\eta}$ and $\eta=\pm 1$, leads to the same solution as in \eq{1D_2by2_edge_xi} with the replacement $A\rightarrow A_{x}$ and $B\rightarrow B_{x}$.
Demanding $\xi_{+}^{-1}+\xi_{-}^{-1}=-\eta A_{x}/B_{x}>0$ yields $\eta=-\sgn(A_{x}/B_{x})$, so the eigenvalue of $\chi_{\eta}$ is determined by the sign of $A_{x}/B_{x}$.
At $M\rightarrow M_\rmc$, one has $\xi_{-}>\xi_{+}$, so the decay length is 
\begin{align}
\begin{split}
\xi_{-}\approx -\sgn\left[B_{x}(M-M_\rmc)\right]\left|\frac{A_{x}}{M-M_\rmc}\right|\;.
\end{split}
\end{align} 
Hence, the zero-energy edge state, which only appears in the band-inverted regime~\cite{Qi11}, where $B_{x}(M-M_\rmc)<0$, has a decay length that coincides with the correlation length in \eq{2by2_Dirac_F0_xi} up to a multiplicative constant that depends on the parameters of the Dirac Hamiltonian.
Note that for the simple models with $A_{1y}=A_{2x}=0$, such as the simple model of a Chern insulator in Sec.~\ref{2by2_Dirac}, the two length scales coincide $\xi_{-}\approx\sqrt{2/3}\;\xi_{x}$ as $M\rightarrow M_\rmc$.
The edge state localized at the interface between a topologically trivial and nontrivial state for this 2D case can be calculated in the same fashion as in Eqs.~\eqref{interface_wave_function} to \eqref{interface_xi_case2}, and the same conclusion is obtained.

To address the edge state in the higher orbital symmetry Dirac model discussed in Sec.~\ref{Universality_class}, we consider Eq.~\eqref{eq:2by2_Dirac_general_p} defined in half-space $x>0$ such that $\phi=0$, and solve the zero energy edge state in real space that satisfies 
\begin{eqnarray}
(\bd\cdot\bsigma)\psi=\left[\eta_{p}\sigma_{x}\left(-\i\right)^{p}\partial_{x}^{p}+M\sigma_{z}\right]\psi=0\;.
\end{eqnarray}
Multiplying the equation by $\sigma_{x}$ and using the wave function $\psi=\chi_{\eta}\phi(x)$, where $\chi_{\eta}$ satisfies $\sigma_{y}\chi_{\eta}=\eta\chi_{\eta}=\pm 1$ and $\phi(x)\propto e^{-x/\xi}$, lead to 
\begin{eqnarray}
\xi^{p}=\frac{\eta_{p}\;\i^{p-1}}{\eta\; M}\;.
\end{eqnarray}
The decay length is identifiable with the real part of one of the roots $\xi_{loc}={\rm Re}(\xi)$, in which the eigen value $\eta$ is chosen such that $\xi_{loc}>0$. Comparing with Eq.~\eqref{xi_scaling}, the coincidence between correlation length and the edge state decay length is evident.

\bibliography{Literatur}

\end{document}